\pdfoutput=1
\documentclass[%
 reprint,
 amsmath,amssymb,
 aps,
 prb,
]{revtex4-2}

\usepackage{graphicx}
\usepackage{subfigure}
\usepackage{url}
\usepackage{dcolumn}
\usepackage{bm}
\usepackage[]{natbib}

\begin{document}

\preprint{APS/123-QED}

\title{Flat bands and magnetism in $\bm{\mathrm{Fe_4 Ge Te_2}}$ and $\bm{\mathrm{Fe_5GeTe_2}}$ due to bipartite crystal lattices}

\author{Fuyi Wang}
\author{Haijun Zhang$^*$}
\affiliation{National Laboratory of Solid State Microstructures, School of Physics, Nanjing University, Nanjing 210093, China.}%
\affiliation{Collaborative Innovation Center of Advanced Microstructures, Nanjing University, Nanjing 210093, China.}


\date{\today}

\begin{abstract}

$\mathrm{Fe_{n=4,5}GeTe_2}$ exhibits quasi-two-dimensional properties as a promising candidate for a near-room-temperature ferromagnet, which has attracted great interest. In this work, we notice that the crystal lattice of $\mathrm{Fe_{n=4,5}GeTe_2}$ can be approximately regarded as being stacked by three bipartite crystal lattices. By combining the model Hamiltonians of bipartite crystal lattices and first-principles calculations, we investigate the electronic structure and the magnetism of $\mathrm{Fe_{n=4,5}GeTe_2}$. We conclude that flat bands near the Fermi level originate from the bipartite crystal lattices and that these flat bands are expected to lead to the itinerant ferromagnetism in $\mathrm{Fe_{n=4,5}GeTe_2}$. Interestingly, we also find that the magnetic moment of the Fe5 atom in $\mathrm{Fe_5 Ge Te_2}$ is distinct from the other Fe atoms and is sensitive to the Coulomb interaction $U$ and external pressure. These findings may be helpful to understand the exotic magnetic behavior of $\mathrm{Fe_{n=4,5} Ge Te_2}$.
\end{abstract}

\maketitle


\section{Introduction}
In recent years, $\mathrm{Fe_{n=3,4,5}GeTe_2}$ have been discovered as van der Waals (vdW) itinerant ferromagnets with a high Curie temperature ${T_c}$ \cite{Fe4m1, Fe5m1, Fe5m2, Fe4, Fe5, Fe4n, Fe5y, Feaf, Fe5pt, Fe2dm}. They are promising for spintronic applications due to the near-room-temperature ferromagnetism, magnetic anisotropy, and high electric conductivity \cite{Fel345, Fe4, Fe5, Fe4n, Fe5y, Feaf, Fe5pt, Fe2dm, Fe42, Fe52, Fe53, Fe5m2, Fe5m5, Felm, Fegm, Fe2mm, 2Dm1, 2Dm2, 2Dm3, 2Dm4, 2Dm5}.
$\mathrm{Fe_{3}GeTe_2}$ was first discovered in $\mathrm{Fe_{n=3,4,5}GeTe_2}$ family\cite{Fe3N, Fe3NM}, and its magnetism and electronic structure have been extensively investigated\cite{Fe3T1, Fe3T2, Fe3IM}. $\mathrm{Fe_{4}GeTe_2}$ and $\mathrm{Fe_{5}GeTe_2}$, with the space groups of $R\bar{3} m$ and $ R3m$, share a similar lattice structure which is distinct from that of $\mathrm{Fe_{3}GeTe_2}$, and especially $\mathrm{Fe_{5}GeTe_2}$ can be considered to be obtained from inserting one Fe layer into $\mathrm{Fe_{4}GeTe_2}$. Therefore, we mainly focus on $\mathrm{Fe_{n=4,5}GeTe_2}$ in this work.
$\mathrm{Fe_{n=4,5}GeTe_2}$ also exhibit many interesting properties, such as the Kondo effect \cite{Fe5k}, anomalous Hall effect (AHE) \cite{Fe5a1, Femo}, butterfly-shaped magnetoresistance \cite{Fe5bm}, controllable topological magnetic transformations \cite{Fe5m4}, and skyrmionic spin structures up to the room temperature \cite{Fems, Fe5s1}. However, the underlying physics of the magnetic behaviors of $\mathrm{Fe_{n=4,5}GeTe_2}$ has not been well understood.

\begin{figure*}[t]
  \includegraphics[width=\textwidth]{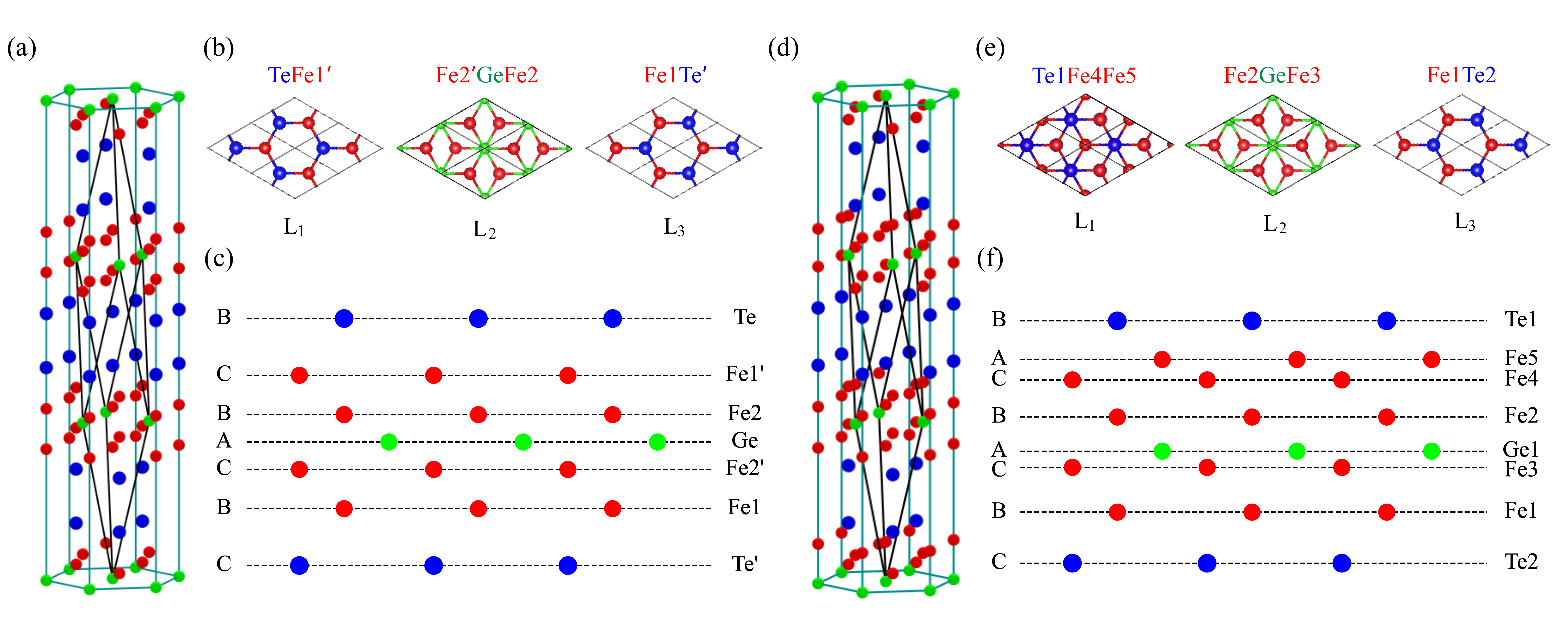}%
  \caption{\label{fig:f1}Crystal structure and bipartite crystal lattices. (a) Crystal structures of $\mathrm{Fe_4GeTe_2}$ with primitive lattice cell in black solid line. (b) Top views of three BCLs of $\mathrm{Fe_4GeTe_2}$, labeled $\mathrm{L_1}$, $\mathrm{L_2}$ and $\mathrm{L_3}$. The first BCL ($\mathrm{L_1}$) is a honeycomb lattice. The center BCL  ($\mathrm{L_2}$) is a dice lattice. The third BCL ($\mathrm{L_3}$) is equivalent to the $\mathrm{L_1}$ due to the inversion symmetry. (c) The side view of the septuple layer of $\mathrm{Fe_4GeTe_2}$. The triangle lattice has three different stacked positions denoted as $A$, $B$ and $C$. (d) Crystal structures of $\mathrm{Fe_5GeTe_2}$ with primitive lattice cell in black solid line. (e) Top view of three  BCLs of $\mathrm{Fe_5GeTe_2}$. The first BCL ($\mathrm{L_1}$) and the center BCL ($\mathrm{L_2}$) are dice lattices. The third BCL ($\mathrm{L_3}$) is a honeycomb lattice. (f) The side view of the octuple layer of $\mathrm{Fe_5GeTe_2}$.}
\end{figure*}

To investigate the electronic structure and the magnetism of $\mathrm{Fe_{n=4,5}GeTe_2}$, we approximately decompose their crystal
lattices into three basic layers due to the layered structure. Each basic layer contains at least one Fe layer and one Ge or Te layer, as shown in Fig.~\ref{fig:f1}. Interestingly, we notice that these basic layers of $\mathrm{Fe_{n=4,5}GeTe_2}$ can be approximately regarded as bipartite crystal lattices (BCLs) which can be divided into two sublattices with negligible intra-sublattice hopping \cite{dice0, FB1}, since the hopping primarily occurs between the Fe and Ge/Te sublattices. We determine that the stacked BCLs in $\mathrm{Fe_{n=4,5}GeTe_2}$ give rise to flat bands \cite{Fe45, Fe4b, Fe5b, Fe5b2} which may account for the ferromagnetism observed in these materials. It is worth mentioning that the decomposition of BCLs for $\mathrm{Fe_{n=4,5}GeTe_2}$ is a rough approximation, and the hoppings between adjacent BCLs still require careful consideration.

In this work, we construct model Hamiltonians for the stacked BCLs of $\mathrm{Fe_{n=4,5} Ge Te_2}$. We determine that the flat bands can be attributed to the BCLs, based on these model Hamiltonians. We also demonstrate that the itinerant ferromagnetism in these materials arises from the nearly flat bands near the Fermi energy driven by the Coulomb interaction $U$, which is known as flat-band ferromagnetism \cite{FBL, FBM, FBM1, FBM2, FBM4, FBM5, FB2F}. The BCL-induced ferromagnetism primarily depends on the lattice structure, orbitals, and electron filling number. We expect that this conclusion could be extended to other vdW ferromagnets. Furthermore, by combining the model Hamiltonians and first-principles calculations, we find that the magnetic moment of Fe5 in $\mathrm{Fe_5 Ge Te_2}$ is sensitive to both ${U}$ and external pressures. The pressure-tunable magnetic moment transitions in $\mathrm{Fe_5 Ge Te_2}$ might be experimentally observed.

\section{Methods}
First-principles calculations are carried out using the Perdew-Burke-Ernzerhof-type (PBE) generalized gradient approximation (GGA) \cite{vasp4} of the density functional theory (DFT), using the Vienna ab initio simulation package (VASP) \cite{vasp1, vasp2, vasp3}. We take the $\mathrm{GGA+U}$ method with $U=3.0\ eV$ to investigate the correlation effects. We calculated the total energy for different magnetic states and determined the ferromagnetic ground state of $\mathrm{Fe_{n=4,5} GeTe_2}$. 
A kinetic energy cutoff of $500\ eV$ is used, and the $10\times 10 \times 10$ k-point mesh is taken for the bulk calculations. The experimental lattice constants of $\mathrm{Fe_4 Ge Te_2}$ ($a = 4.03\ \mathrm{\AA}$, and $c = 29.08\ \mathrm{\AA}$) \cite{Fe4} and $\mathrm{Fe_5 Ge Te_2}$ ($a = 4.04\ \mathrm{\AA}$, and $c = 29.19\ \mathrm{\AA}$)\cite{Fe5} are adopted. The inner atomic positions are obtained via full relaxation with a total energy tolerance of ${10^{-6}}$ eV. The Wannier-based model Hamiltonians are obtained from the projection of the $p$ orbitals of Ge and Te and the $d$ orbitals of Fe through employing the WANNIER90 package \cite{W90, W902, W903}.  

\section{Crystal structure and orbitals}
\subsection{The bipartite crystal lattice }

$\mathrm{Fe_4GeTe_2}$ has a space group of ${R{\bar{3}}m (166)}$ that includes an inversion symmetry. The Te (Fe1$'$, Fe2) atoms are related to the Te$'$ (Fe1, Fe2$'$) atoms via an inversion operation, with the Ge atom serving as the inversion center. The lattice of $\mathrm{Fe_4GeTe_2}$ has a septuple-layer structure, as shown in Fig.~\ref{fig:f1}. When we use ABC to represent different stacking positions, the stacking manner of $\mathrm{Fe_4 GeTe_2}$ in the septuple-layer structure can be expressed as Te(B)-Fe1$'$(C)-Fe2(B)-Ge(A)-Fe2(C)-Fe1$'$(B)-Te$'$(C), as shown in Fig.~\ref{fig:f1}(c). We can decompose the unit cell of $\mathrm{Fe_4GeTe_2}$ into three BCLs stacked along the $z$ direction, denoted as $\mathrm{L_1}$, $\mathrm{L_2}$, and $\mathrm{L_3}$. The $\mathrm{L_1}$ BCL comprises the Fe1 sublattice and the Te sublattice, while the $\mathrm{L_3}$ BCL comprises the Fe1$'$ sublattice and the Te$'$ sublattice. The $\mathrm{L_1}$ and $\mathrm{L_3}$ BCLs are related through the inversion symmetry. The $\mathrm{L_2}$ BCL consists of the Fe2(Fe2$'$) sublattice and the Ge sublattice, which can be viewed as a dice lattice \cite{dice0, FB2, ar1, ar2} with an inversion symmetry, where the Ge atom acts as the inversion center.

$\mathrm{Fe_5 Ge Te_2}$ belongs to the ${R3m (160)}$ space group. The lattice of $\mathrm{Fe_5 Ge Te_2}$ can be obtained by inserting a Fe5 layer between the Fe1$'$ and Te1 layers of $\mathrm{Fe_4GeTe_2}$ \cite{Fe4}, as depicted in Fig.~\ref{fig:f1}(c, f). The inserted Fe5 layer breaks the inversion symmetry, resulting in the unequivalence of Fe1 and Fe1$'$, as well as Fe2 and Fe2$'$. Consequently, Fe1$'$ and Fe2$'$ in $\mathrm{Fe_5GeTe_2}$ are renamed as Fe4 and Fe3, respectively. The lattice of $\mathrm{Fe_5GeTe_2}$ has an octuple-layer structure, which can also be approximately decomposed into three BCLs stacked along the $z$ direction. The stacking manner of $\mathrm{Fe_5 GeTe_2}$ in the octuple-layer structure can be expressed as Te1(B)-Fe5(A)-Fe4(C)-Fe2(B)-Ge(A)-Fe3(C)-Fe1(B)-Te2(C), as shown in Fig.~\ref{fig:f1}(f). The $\mathrm{L_1}$ BCL in $\mathrm{Fe_5GeTe_2}$ is a dice lattice consisting of the Fe4 and Fe5 sublattice and the Te1 sublattice, though it is just a quasi-BCL due to the non-negligible nearest hopping between Fe4 and Fe5. The $\mathrm{L_2}$ BCL in $\mathrm{Fe_5GeTe_2}$ is a dice lattice but lacks the inversion symmetry. Lastly, the $\mathrm{L_3}$ BCL in $\mathrm{Fe_5GeTe_2}$ is a honeycomb lattice which is almost identical to the $\mathrm{L_3}$ BCL in $\mathrm{Fe_4GeTe_2}$.

\subsection{Orbitals and the site symmetry}
The ferromagnetism in $\mathrm{Fe_{n=4,5}GeTe_2}$ is mainly due to the partially filled $d$ orbitals of Fe, whereas Te and Ge do not exhibit major magnetic behavior. Therefore, it is important to analyze the splitting of Fe's $d$ orbitals to reveal
the underlying mechanism of the ferromagnetism.

Both $\mathrm{Fe_4GeTe_2}$ and $\mathrm{Fe_5GeTe_2}$ have the $C_{3v}$ site symmetry, which has three irreparable representations: two 1D irreps ${A_{1,2}}$ and a 2D irrep ${E}$. The ${p}$ and ${d}$ orbitals can be classified based on the irreducible representations (irrep) of the site symmetry group. Here, ${p_z}$ corresponds to ${A_1}$ irrep for ${p}$ orbitals of Te and Ge, while (${p_x}$, ${p_y}$) corresponds to ${E}$ irrep and forms a doublet. Moreover, the $d$ orbitals of Fe are divided into a singlet ${d_{z^2}}$ with ${A_1}$ irrep and two doublets (${d_{xz}}$, ${d_{yz}}$) and (${d_{xy}}$, ${d_{x^2-y^2}}$) with ${E}$ irrep. The orbitals in doublets can be recombined as ${p_x \pm i p_y}$, ${d_{xz} \pm i d_{yz}}$, and ${d_{x^2-y^2} \pm i d_{xy}}$, renamed according to their quantum numbers of the angular momentum projection operator $\hat{l_z}$ as ${p_{m=\pm1}}$, ${d_{m=\pm1}}$, and ${d_{m=\pm2}}$. Meanwhile, the singlet orbitals are renamed as ${p_{m=0}}$ and ${d_{m=0}}$, respectively. 

Due to the stacking of BCLs along the $z$-direction, the most significant interactions between adjacent layers are $\sigma$ bondings formed between the $p_z$ and $d_{z^2}$ orbitals, both of which have the quantum numbers $m=0$. 

\begin{table}[htbp]
  \caption{\label{tab:table1}
  The character table of the point group ${C_{3v}}$ 
  }
  \begin{ruledtabular}
  \begin{tabular}{cccccc}
  \textrm{}&
  \textrm{${E}$}&
  \textrm{${C_3 (z)}$}&
  \textrm{${3 \sigma_z}$}&
  \textrm{${p}$ orbitals}&
  \textrm{${d}$ orbitals}\\
  \colrule
  ${A_1}$ & 1 & 1 & 1 & ${p_z}$ & ${d_{z^2}}$\\
  ${A_2}$ & 1 & 1 &-1 & / & / \\
  ${E}$   & 2 &-1 & 0 & (${p_x,p_y}$)& (${d_{xy},d_{x^2-y^2}}$); (${d_{xz},d_{yz}}$)
  \end{tabular}
  \end{ruledtabular}
\end{table}

\begin{figure}[t]
  \centering
  \includegraphics[width=0.48\textwidth]{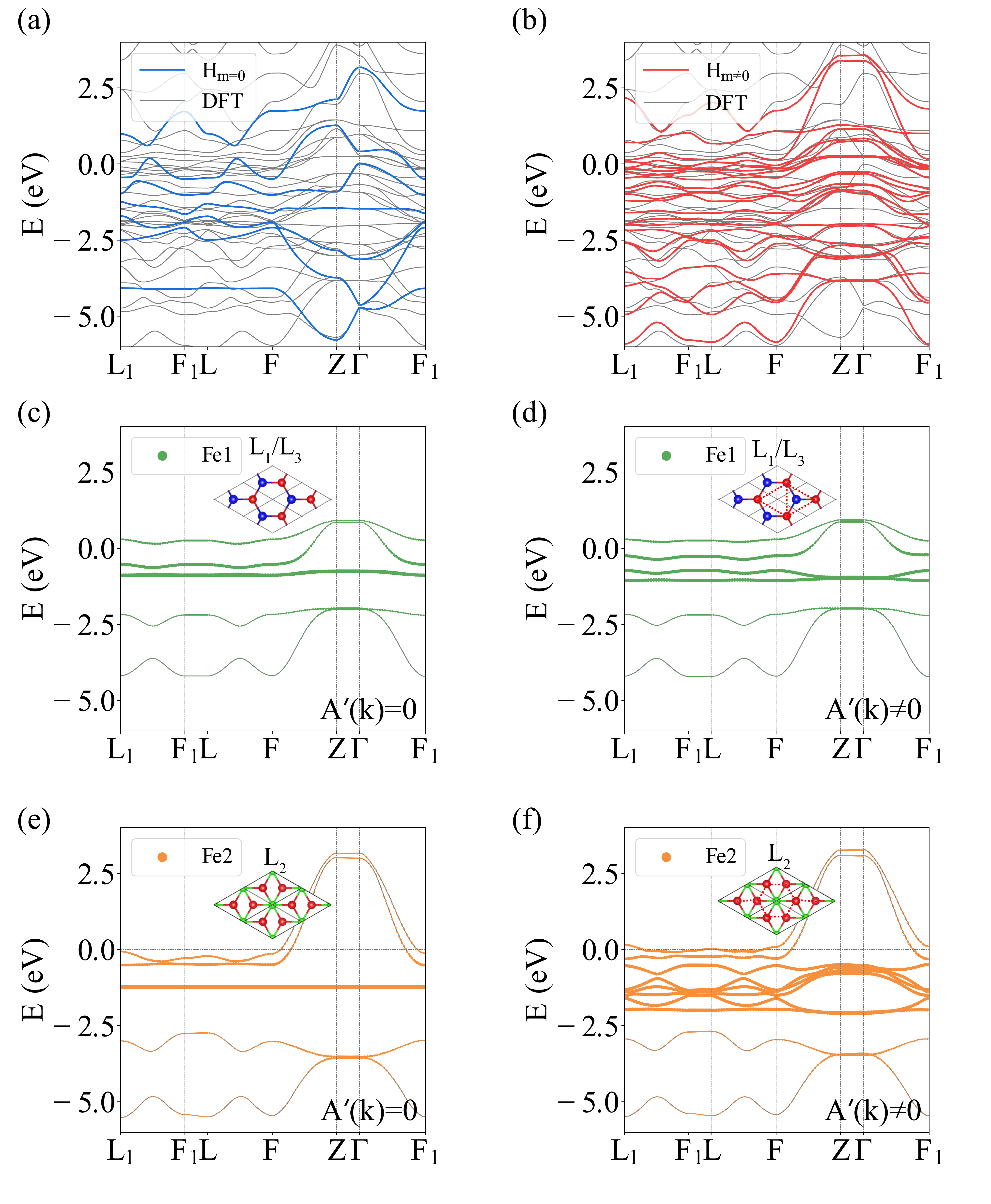}
  \caption{\label{fig:f2} Band structures by model Hamiltonians of $\mathrm{Fe_4GeTe_2}$. (a, b) The band structures of $H_{m=0}$ and ${H_{m \ne 0}}$ in red and blue. The dashed gray lines are the band structures calculated by first-principles calculations (density functional theory, DFT). (c, d) Band structures by the $H_{B_1}/H_{B_3}$ for $A'(k)=0$ (c) and $A'(k) \neq 0$ (d). $H_{B_1}$ is equivalent to $H_{B_3}$ due to the inversion symmetry. The projections of Fe1 are in green. (e, f) Band structures by the $H_{B_2}$ for $A'(k)=0$ (e) and $A'(k) \neq 0$ (f). The projections of Fe1 are in orange.}
\end{figure}

\section{Model Hamiltonians} 

\subsection{Construction of Model Hamiltonians}

To understand the origin of flat bands, it is essential to formulate a model Hamiltonian. Since the lattice structure of $\mathrm{Fe_{n=4,5}GeTe_2}$ can be viewed as three stacked BCLs, we construct the tight-binding model Hamiltonian $H_{tot}$ by placing the BCL Hamiltonians ${H_{L_i}}$ on the diagonal. The general form of the tight-binding model Hamiltonian for $\mathrm{Fe_{n=4,5}GeTe_2}$ is written as, 

\begin{equation}
  H_{tot}(k)=\left(
    \begin{array}{ccc}
      H_{L_1}(k) &S_{12}(k) &S_{13}(k)\\
      S_{12}^\dagger(k) &H_{L_2}(k) &S_{23}(k)\\
      S_{13}^\dagger(k) &S^\dagger_{23}(k) &H_{L_3}(k)
    \end{array}
  \right)
\end{equation}

\begin{figure}[t]
  \centering
  \includegraphics[width=0.48\textwidth]{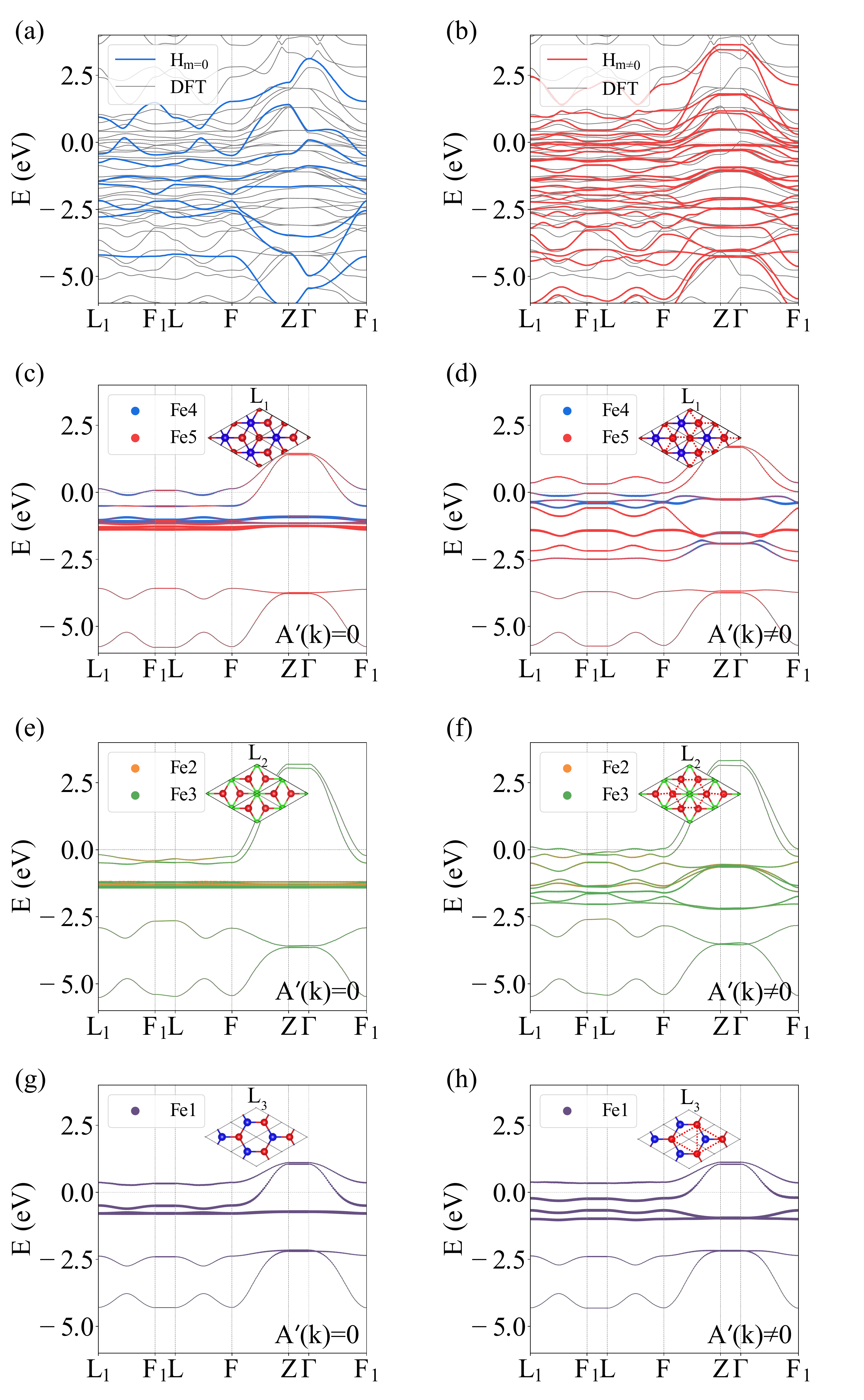}
  \caption{\label{fig:f3} Band structures by model Hamiltonians of $\mathrm{Fe_5GeTe_2}$.  (a, b) The band structures of $H_{m=0}$ and ${H_{m \ne 0}}$ in red and blue. The dashed gray lines are the band structures calculated by first-principles calculations. (c, d) Band structures by the $H_{B_1}$ for $A'(k)=0$  (c) and $A'(k) \neq 0$ (d). The projections of Fe4 and Fe5 are in blue and red, respectively. (e, f) Band structures by the $H_{B_2}$ for $A'(k)=0$ (e) and $A'(k) \neq 0$ (f). The projections of Fe4 and Fe5 are in orange and green, respectively.  (g, h) Band structures by the $H_{B_3}$ for $A'(k)=0$ (g) and $A'(k) \neq 0$ (h). The projections of Fe1 are in purple. }
\end{figure}

where the ${S_{12}}$ and ${S_{23}}$ represent the hopping between adjacent BCLs which usually have the same order of magnitude as the intra-BCL hoppings, while the ${S_{13}}$ between the L$_1$ and L$_3$ BCLs is almost zero. Therefore, the Hamiltonians of $\mathrm{L_1}$, $\mathrm{L_2}$, and $\mathrm{L_3}$ BCLs cannot be treated independently. However, the hopping between adjacent BCLs primarily occurs between orbitals along the ${z}$ direction, such as the ${p_z}$ and ${d_{z^2}}$ with ${m=0}$. Therefore, to simplify the model, the orbitals can be categorized into two sets. The first set comprises all the orbitals with ${m=0}$, while the second set consists of the remaining orbitals with ${m\neq 0}$. By applying a unitary transformation, the original model Hamiltonian is transformed to,

\begin{equation}
  H_{tot}(k)=\left(
    \begin{array}{cc}
      H_{m=0}(k) &S_m(k)\\
      S_m^\dagger(k) &H_{m\neq 0}(k)
    \end{array}
  \right)
\end{equation}
where the ${H_{m=0}(k)}$ and ${H_{m\neq 0}(k)}$ are the Hamiltonian with the ${m=0}$ orbitals and ${m\neq 0}$ orbitals respectively. ${S_m(k)}$ is the hopping matrix between the ${m=0}$ and ${m\neq 0}$ orbitals. Fig.~\ref{fig:f2}(b) and Fig.~\ref{fig:f3}(b) show that the band structures calculated by ${H_{m \neq 0}(k)}$ of $\mathrm{Fe_{n=4,5} Ge Te_2}$ can catch the main feature of the band structures from the first-principles calculations, which validate the partitioning of orbitals into ${m=0}$ and ${m \neq 0}$. Since the strongest interlayer coupling occurs among the orbitals with $m=0$, we treat ${H_{m=0}(k)}$ as a whole without decomposition. 

Since the hopping between adjacent BCLs is relatively weak for the ${m\neq 0}$ orbitals, the ${H_{m\neq 0}(k)}$ is given by, 
\begin{equation}
  H_{m\neq 0}(k)=\left(
    \begin{array}{ccc}
      H_{B_1}^{m\neq 0}(k) &S_{B_{12}}(k) &0\\
      S_{B_{12}}^\dagger(k) &H_{B_2}^{m\neq 0}(k) &S_{B_{23}}(k)\\
      0 &S^\dagger_{B_{23}}(k) &H_{B_3}^{m\neq 0}(k)
    \end{array}
  \right)
\end{equation}
where ${H^{m\neq 0}_{B_i}}$ is the Hamiltonian based on the ${m\neq 0}$ orbitals of the L$_i$ BCL, and ${S_{B_{ij}}}$ is the hopping between the L$_{i}$ BCL and the L$_{j}$ BCL, which can be negligible (${S_{B_{ij}}}\approx0$). Then, ${H_{m\neq 0}(k)}$ is further considered to made up of the three individual ${H^{m\neq 0}_{B_i}}$ which is written as \cite{FB1},

\begin{equation}
  H_{B_i}^{m\neq 0}(k)=\left(
    \begin{array}{cc}
      A(k) &S(k)\\
      S^\dagger(k) &B(k)
    \end{array}
  \right)
\end{equation}
where ${A(k)}$/${B(k)}$ is a Hermitian matrix denoting the onsite energy and intra-sublattice hopping and ${S(k)}$ denotes the inter-sublattice hopping for each BCL. Since onsite energies lie on the diagonal, the matrix $A'(k)/B'(k)$ obtained after removing the diagonal terms of $A(k)/B(k)$ represents the intra-sublattice hoppings. As mentioned above, the decomposition of BCLs of $\mathrm{Fe_{n=4,5}GeTe_2}$ is a rough approximation due to the existence of the nonzero intra-sublattice hoppings which leads to the dispersion of flat bands. 

\subsection{Flat bands due to BCLs}

In general, a BCL Hamiltonian can induce ($N=N_A-N_B$) flat bands when ${N_A>N_B}$ \cite{FB1}, as shown in Appendix A. Here, $N_{A/B}$ denotes the number of orbitals present on the $A/B$ sublattice. The emergence of flat bands can be attributed to the destructive interference of wavefunctions associated with the properties of the BCL \cite{FB3, FBk}. When ${N_A>N_B}$, the hopping along different directions overlaps destructively at the $B$ sublattice, resulting in (${N_A-N_B}$) states solely on the $A$ sublattice at every momentum $k$. Since there are no hoppings between states on the $A$ sublattice of BCL, these states form (${N_A-N_B}$) flat bands. However, according to the proof in Appendix A, the crystal field splitting of orbitals due to the crystal field effect and the intra-sublattice hoppings on $A$ sublattice may lead to slight bending or loss of degeneracy in the flat bands. We will take into account the impact of the crystal field splitting and $A'(k)$ on the flat bands when performing calculations using these model Hamiltonians. 

We first analyze the flat bands according to the BCL Hamiltonian ${H_{B_i}}$ of $\mathrm{Fe_{n=4,5} Ge Te_2}$. We first neglect the crystal field splitting of the $d$ orbital due to the crystal field effect and the intra-sublattice hoppings $A'(k)$ for each BCL. The $\mathrm{L_1}$/$\mathrm{L_3}$ BCL of $\mathrm{Fe_4 Ge Te_2}$ and $\mathrm{L_3}$ BCL of $\mathrm{Fe_5 Ge Te_2}$ are honeycomb lattices that consist of a Fe sublattice (denoted as the $A$ sublattice) and a Te sublattice (denoted as the $B$ sublattice). The $A$ sublattice comprises four degenerate $d$ orbitals, while the $B$ sublattice comprises two degenerate $p$ orbitals. As a result, the BCL Hamiltonian with (${N_A-N_B=2}$) degenerate flat bands. On the other hand, the $\mathrm{L_2}$ BCL of $\mathrm{Fe_4GeTe_2}$ and $\mathrm{L_1}$/$\mathrm{L_2}$ BCLs of $\mathrm{Fe_5GeTe_2}$ are dice lattices that consist of the sublattice with two Fe (denoted as the $A$ sublattice) and a Te/Ge sublattice (denoted as the $B$ sublattice). The $A$ sublattice contains eight degenerate $d$ orbitals, while the $B$ sublattice contains two degenerate $p$ orbitals. The BCL Hamiltonian for this dice lattice has (${N_A-N_B=6}$) degenerate flat bands. Due to the relatively localized nature of the $d$ orbitals, we anticipate that the intra-sublattice hopping ($A'(k)$) will have small magnitudes, thereby having a limited impact on the formation of flat bands.

\begin{figure}[t]
  \centering
	\includegraphics[width=0.45\textwidth]{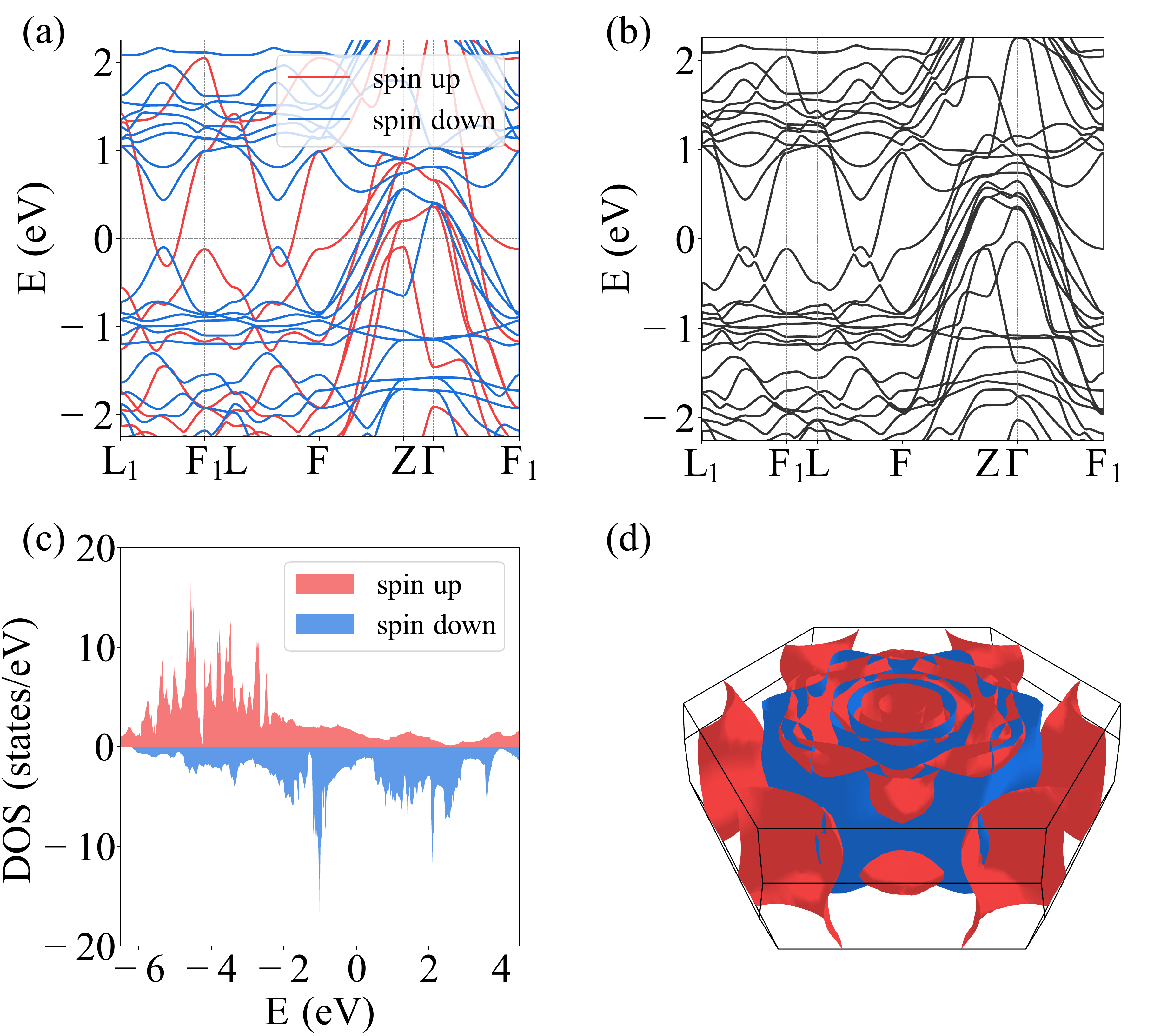}
  \caption{\label{fig:f4} Electronic structures of $\mathrm{Fe_4 Ge Te_2}$ calculated by first-principles calculations with ${U=3.0\ eV}$. (a) The non-SOC band structure. (b) The SOC band structure. (c) The spin-polarized DOS without SOC. (d) The Fermi surfaces with up spin are in red, while the Fermi surfaces with down spin are in blue.}
\end{figure}

Based on the model Hamiltonians, the flat bands are calculated, shown in Fig.~\ref{fig:f2}(c, e) and Fig.~\ref{fig:f3}(c, e, g) without considering the intra-sublattice hoppings in $A$ sublattice ($A'(k)=0$) and Fig.~\ref{fig:f2}(d, f) and Fig.~\ref{fig:f3}(d, f, h) with considering the intra-sublattice hoppings($A'(k)\neq 0$). The clear flat bands have been shown in Fig.~\ref{fig:f2} and Fig.~\ref{fig:f3}, though the $A'(k)\neq 0$ lead to the slight bending of the nearly flat bands. We can see that the dispersion of the bands almost keep unchanged with $A'(k)=0$ and $A'(k)\neq 0$ for $\mathrm{L_1/L_3}$ BCLs of $\mathrm{Fe_{4}GeTe_2}$ and $\mathrm{L_3}$ BCL of $\mathrm{Fe_{5}GeTe_2}$, whereas this is not the case for the $\mathrm{L_2}$ BCL of $\mathrm{Fe_{4}GeTe_2}$ and $\mathrm{L_1/L_2}$ BCLs of $\mathrm{Fe_{5}GeTe_2}$ [Fig.~\ref{fig:f2}(e, f), Fig.~\ref{fig:f3}(c, d, e, f)] due to the hopping between orbitals of Fe in dice lattice. We find that the bands from BCL model Hamiltonians with $A'(k) \neq 0$ can well reproduce the bands obtained from first-principles calculations, which support that the flat bands originate from the BCLs of $\mathrm{Fe_{n=4,5}GeTe_2}$.

It is worth discussing whether the flat bands of $\mathrm{Fe_{n=4,5}GeTe_2}$ are itinerant or local. Flat bands can be classified into two types: trivial flat atomic bands and non-trivial flat bands \cite{FB3}. Flat atomic bands originate from the localization of orbitals or isolated atoms, resulting in negligible overlaps between atomic wavefunctions \cite{FB3}. Conversely, non-trivial flat bands emerge from extended wavefunctions with substantial overlaps and hoppings \cite{FB3}, indicating the itinerant character. In the case of $\mathrm{Fe_{n=4,5}GeTe_2}$, the significant overlaps and hoppings between the orbitals suggest that their flat bands are itinerant.

\begin{figure}[t]
  \centering
  \includegraphics[width=0.48\textwidth]{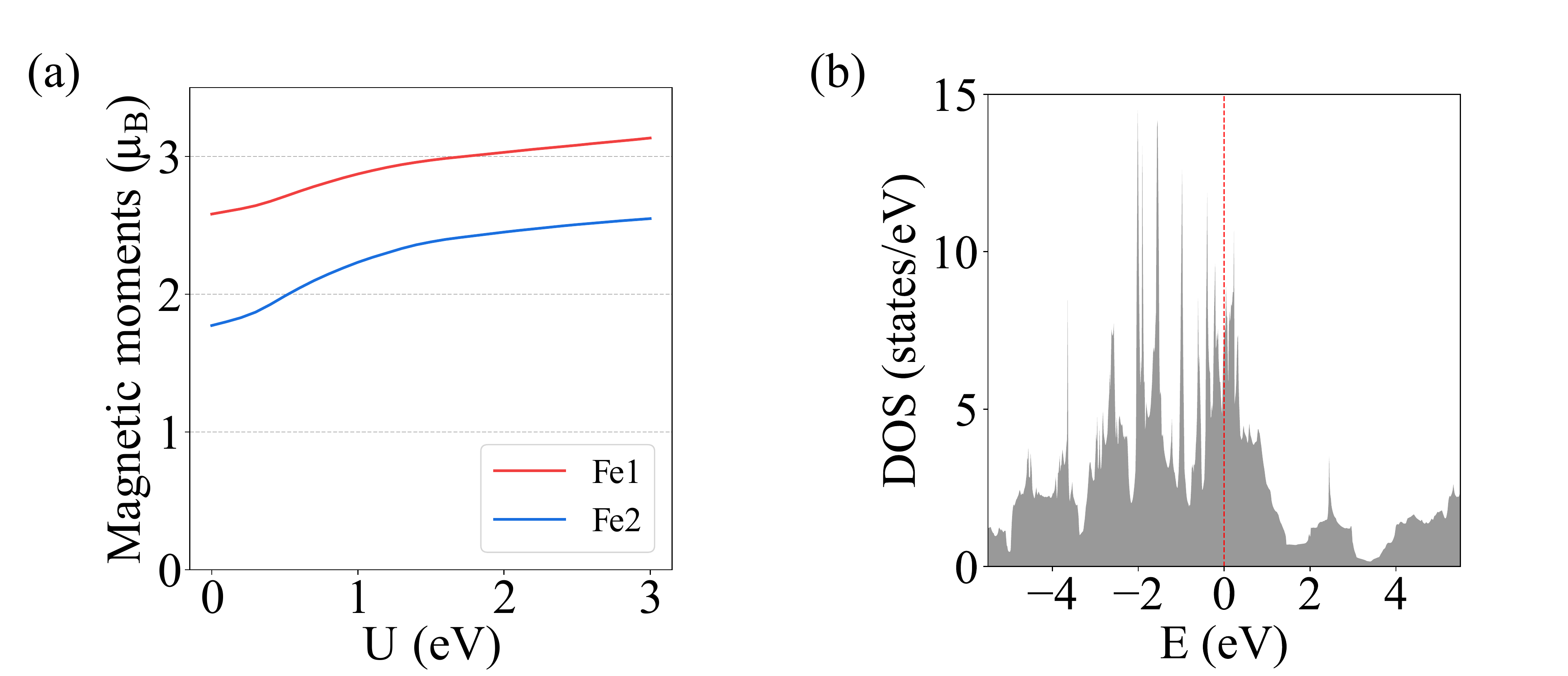}
  \caption{\label{fig:f5} Magnetic properties of $\mathrm{Fe_4 Ge Te_2}$ calculated by first-principles calculations. (a) The ${U}$ dependence of magnetic moments of unequivalent Fe atoms.  (b) The non-spin-polarized DOS. }
\end{figure}

\subsection{Flat-Band Ferromagnetism}
In the absence of spin polarization, all flat bands formed by the $d$ orbitals of Fe are close to the Fermi energy due to the partial occupation of the $d$ orbitals. These flat bands result in sharp peaks of the non-spin-polarized density of states (DOS) near the Fermi energy. According to the Stoner theory, these peaks can lead to spontaneous magnetization \cite{Itm1, Itm2, Itm3}. The critical condition for the instability is expressed as ${U>1/N_{E_F}}$ \cite{Itm1}, and here $N_{E_F}$ denotes the DOS at the Fermi energy. 

As the value of ${U}$ increases, the energies of states with up and down spin will decrease and increase respectively, leading to a spin-polarized DOS. Consequently, flat bands near the Fermi energy in non-spin-polarized band structures can give rise to ferromagnetism. The spin-polarized DOS contributes to the magnetic moment, which can be quantified as ${m=n_{\uparrow}-n_{\downarrow}}$, where $n_{\uparrow}$/$n_{\downarrow}$ represents the number of occupied states with the up/down spin. The magnetic moment increases with increasing ${U}$, which is also confirmed by the results of first-principles calculations [Fig.~\ref{fig:f5}(a) and Fig.~\ref{fig:f7}(a)].

\section{Electronic structure and magnetic properties}

\subsection{$\mathrm{Fe_{4}GeTe_2}$}

We perform first-principles calculations to investigate the electronic structure and magnetic properties of $\mathrm{Fe_4GeTe_2}$. In our calculations, we employ ${U=3.0\ eV}$ to obtain the band structure, DOS, and Fermi surfaces. The results suggest that the band structures with and without SOC are similar, implying that SOC has a negligible effect on the electronic structure of $\mathrm{Fe_4GeTe_2}$ [Fig.~\ref{fig:f4}(a,b)]. 	The non-spin-polarized DOS indicates the $N_{E_F}$ value is about $8.8\ states/eV$. Considering the Stoner criterion as $U>1/N_{E_F}$, $\mathrm{Fe_4 GeTe_2}$ satisfying the Stoner criterion requires U to be greater than $0.11\ eV$. 
Although the precise value of U cannot be ascertained, according to this paper\cite{Fel345}, the effective U value in $\mathrm{Fe_4 GeTe_2}$ is significantly higher than $0.11\ eV$. Hence, our calculations indicate that $\mathrm{Fe_4 GeTe_2}$ should meet the Stoner criterion.
The spin-polarization DOS is consistent with the ferromagnetism [Fig.~\ref{fig:f4}(c)]. The band structure and Fermi surfaces [Fig.~\ref{fig:f4}(d)] indicate the $\mathrm{Fe_4GeTe_2}$ is a quasi-2D ferromagnetic metal. 

\begin{figure}[t]
  \centering
  \includegraphics[width=0.45\textwidth]{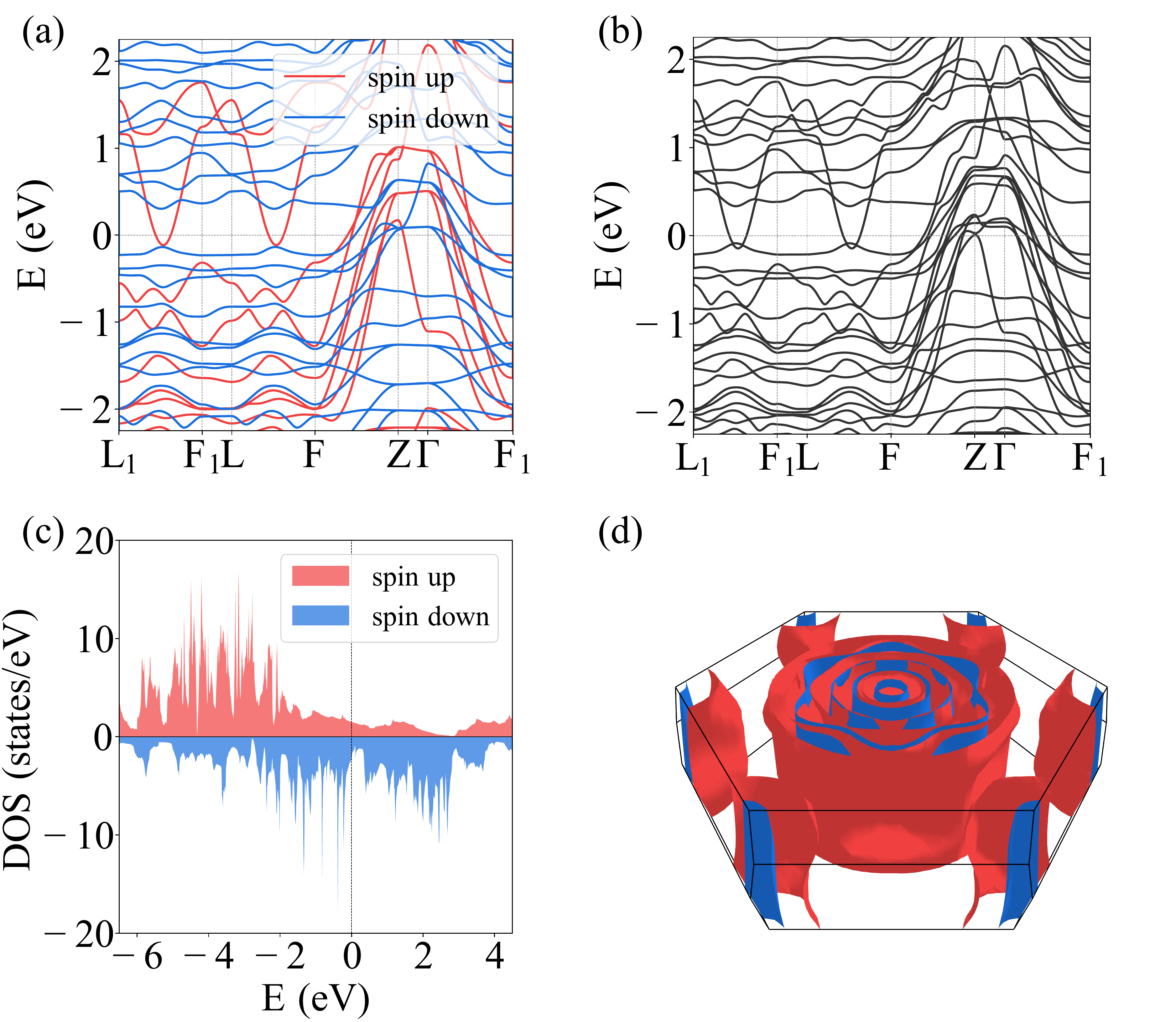}
  \caption{\label{fig:f6} Electronic structures of $\mathrm{Fe_5 Ge Te_2}$ calculated by first-principles calculations with ${U=3.0\ eV}$. (a) The band structure without SOC. (b) The band structure with SOC. (c) The spin-polarized DOS without SOC. (d) The Fermi surfaces with up spin are in red, while the Fermi surfaces with down spin are in blue.}
\end{figure}

By gradually increasing the ${U}$, we observe a gradual increase in the magnetic moments of the Fe atoms of $\mathrm{Fe_4GeTe_2}$.  As illustrated in Fig.~\ref{fig:f5}(a), the magnetic moments of Fe1 (Fe1$'$) and Fe2 (Fe2$'$) surpass ${1.5\ \mu_B}$ when ${U=0.0\ eV}$. Furthermore, we identify the presence of nearly flat bands in the non-spin-polarized band structures  [Fig.~\ref{fig:f2}] and the corresponding peaks in DOS [Fig.~\ref{fig:f5}(b)]. These sharp peaks near the Fermi energy suggest that $\mathrm{Fe_4GeTe_2}$ exhibits characteristics of an itinerant flat-band ferromagnet \cite{Itm1}.

\subsection{$\mathrm{Fe_{5}GeTe_2}$}

\begin{figure}[t]
  \centering
  \includegraphics[width=0.48\textwidth]{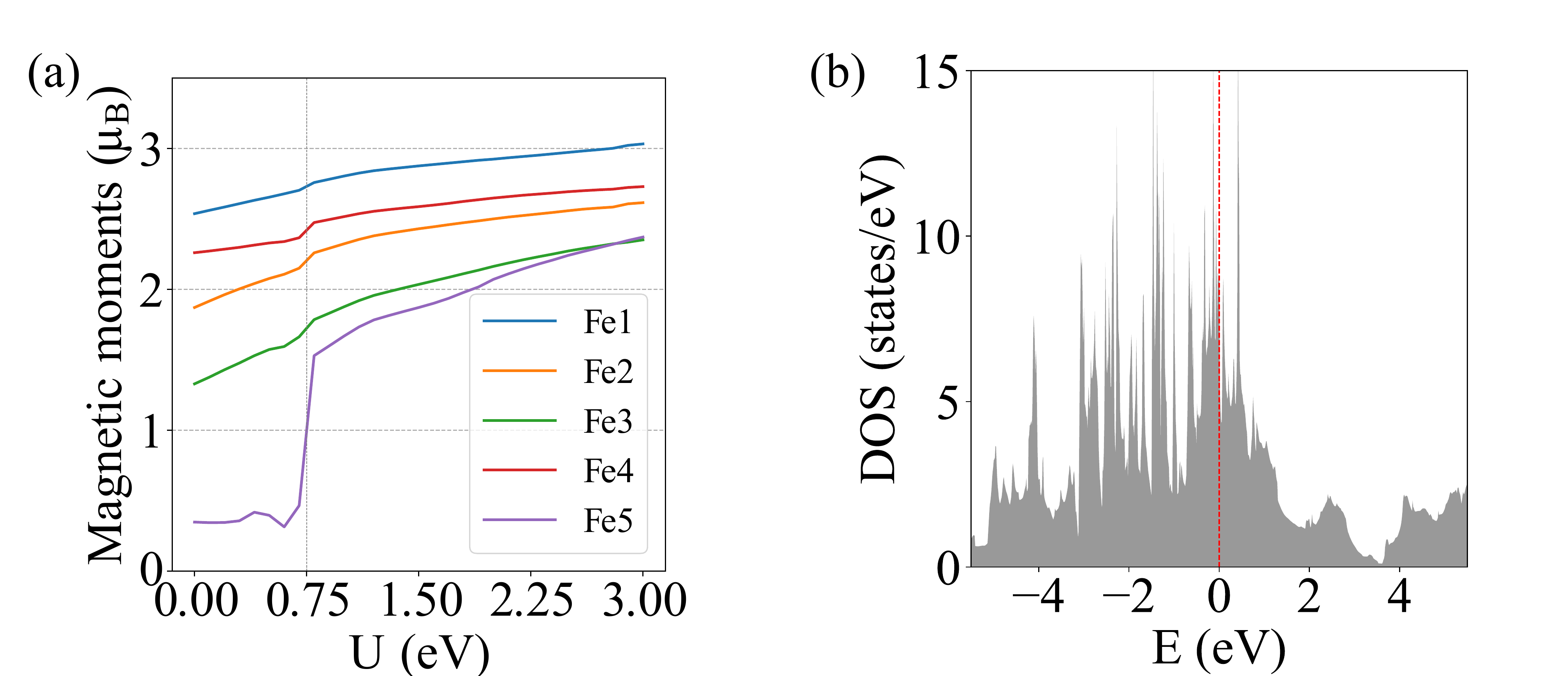}
  \caption{\label{fig:f7} Magnetic properties of $\mathrm{Fe_5 Ge Te_2}$ calculated by first-principles calculations. (a) The ${U}$ dependence of magnetic moments of unequivalent Fe atoms.  (b) The non-spin-polarized DOS. }
\end{figure}

We further perform first-principles calculations to analyze the electronic and magnetic properties of $\mathrm{Fe_5GeTe_2}$. As shown in Fig.~\ref{fig:f6} and Fig.~\ref{fig:f7}, the band structures, DOS, and Fermi surfaces are similar to those of $\mathrm{Fe_4GeTe_2}$. The non-spin-polarized DOS indicates the $N_{E_F}$ value is about $7.4\ states/eV$. The effective value of U is greater than $0.14\ eV$\cite{Fel345}, which satisfies the Stoner criterion. Therefore $\mathrm{Fe_5 GeTe_2}$ should also be an itinerant ferromagnet.

However, there is a significant difference in the magnetic properties of Fe5. For ${U\leq 0.7\ eV}$, the magnetic moments of Fe5 are negligible, while it has a sudden increase between ${U=0.7\ eV}$ and ${0.8\ eV}$. We explain this phenomenon based on the band structure of the $\mathrm{L_1}$ BCL which is a quasi-dice lattice. The energy levels of Fe5 orbitals are slightly lower than those of Fe4 due to their different coupling to Te1. Then, bonding and anti-bonding bands are formed through the hopping between Fe4 and Fe5 orbitals. The anti-bonding band is primarily composed of Fe4 orbitals, whereas the bonding band is dominated by Fe5 orbitals. The Fe4-dominated bands are very close to the Fermi energy, resulting in the spontaneous magnetization of Fe4. As the value of ${U}$ increases, the Fe5-dominated flat bands cross the Fermi level, leading to a pronounced enhancement in Fe5's magnetic moment.

\begin{figure}[t]
  \centering
  \includegraphics[width=0.48\textwidth]{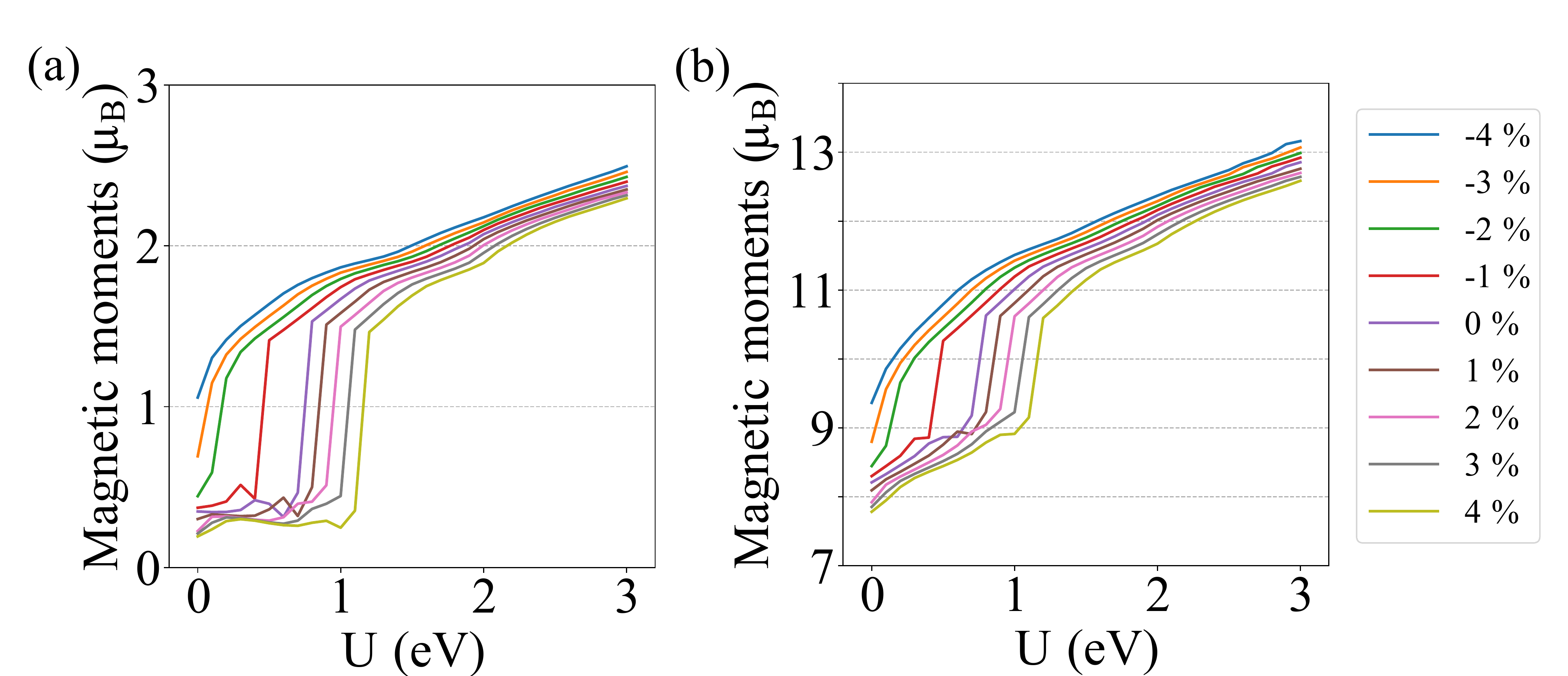}
  \caption{\label{fig:f8} The dependence of the magnetism of $\mathrm{Fe_5GeTe_2}$ on external pressures calculated by first-principles calculations. The percentages indicate the proportion of deformation in the $z$-direction. (a) The ${U}$ dependence of Fe5's magnetic moments under different external pressures. (b) The ${U}$ dependence of total magnetic moments under different external pressures. }
\end{figure}

We also investigate the effect of external pressure on the magnetic moment of Fe5 for  $\mathrm{Fe_5GeTe_2}$. For simplicity, the cell volume is kept unchanged, applying pressure along the ${z}$-direction causes stretching in the ${xy}$ plane. The compression along the ${z}$-direction is primarily accommodated by the vdW gaps, resulting in negligible alteration to the vertical spacing among atoms within each octuple layer. Consequently, the pressure primarily influences the intra-layer hoppings due to the in-plane stretching. Therefore, the hopping between Fe4 and Fe5 slightly decreases, causing the energy level of Fe5 to approach the Fermi energy. Therefore, the magnetic moment of Fe5 increases with the pressure in the ${z}$ direction, as illustrated in Fig.~\ref{fig:f8}.  
Though pressure-induced magnetic transitions of Fe5 may be detectable through neutron scattering\cite{Fe3nd}, such measurements are relatively challenging. Our calculations find that the total magnetic moment changes with the magnetic variation of Fe5, which can be readily quantified. Therefore, measuring the total magnetic moment is a more convenient approach to indicate the magnetic transitions of Fe5.
\section{Conclusion}

In this study, we investigate the origin of the nearly flat bands and ferrimagnetism in $\mathrm{Fe_{n=4,5} Ge Te_2}$. Our analysis reveals that the lattice structure of these materials can be viewed as three stacked BCLs along the ${z}$ direction. The presence of different orbital numbers on two sublattices results in nearly flat bands. We demonstrate that the observed ferromagnetism in $\mathrm{Fe_{n=4,5} Ge Te_2}$ arises from these nearly flat bands according to the Stoner theory. By combining model calculations with first-principles calculations, we find that the magnetic moment of Fe5 in $\mathrm{Fe_5 Ge Te_2}$ is sensitive to Coulomb interactions ${U}$ and external pressure, which might be experimentally observed.

The emergence of flat-band ferromagnetism in $\mathrm{Fe_{n=4,5} Ge Te_2}$ predominantly depends on the lattice structure, orbital characteristics, and electron filling number. These findings contribute to our understanding of the electronic and magnetic properties of vdW ferromagnets, specifically $\mathrm{Fe_{n=4,5} Ge Te_2}$. 

\begin{acknowledgments}

This work is supported by National Key Projects for Research and Development of China (Grant No.2021YFA1400400 and No.2017YFA0303203), the Fundamental Research Funds for the Central Universities (Grant No. 020414380185), Natural Science Foundation of Jiangsu Province (No. BK20200007), the Natural Science Foundation of China (No. 12074181, No. 12104217, and No. 11834006) and the Fok Ying-Tong Education Foundation of China (Grant No. 161006). 

\end{acknowledgments}

\appendix
\section{A brief proof of flat bands in BCLs}
Firstly we ignore the intra-sublattice hoppings and the crystal field splitting of orbitals on the $A$ sublattice, so $A_k = \epsilon I$, where $I$ is an identity matrix and $\epsilon$ is the onsite energies of orbitals on $A$ sublattice.
We set $\epsilon$ as the zero energy point:

\begin{equation}
  H_k = \begin{pmatrix} O & S_k \\ S_k^\dagger & B_k \end{pmatrix}
\end{equation}

Diagonalizing $N_B\times N_A$ rectangular matrix $S_k$, we have:

\begin{equation}
  S_k = W_k \Sigma_k V_k^\dagger
\end{equation}

Here, $\Sigma_k$ is a rectangular diagonal matrix with $N_A - N_B$ zero rows:

\begin{equation}
  \Sigma_k = \begin{pmatrix} \epsilon_1 &0 &0 &\dots &0 \\ 0 &\epsilon_2 &0 &\dots &0 \\ 0 &0 &\epsilon_3 &\ddots &0 \\ \vdots &\ddots &\ddots &\ddots &0 \\ 0 &0 &\dots &0 &\epsilon_{N_B} \\ 0 &0 &\dots &0 &0\\ \vdots &\vdots &\dots &\vdots &\vdots \\ 0 &0 &\dots &0 &0 \end{pmatrix}
\end{equation}

Then we perform a similarity transformation on $H_k$ as:

\begin{equation}
  H_k = \begin{pmatrix} W_k & O \\ O & V_k \end{pmatrix} \begin{pmatrix} O & \Sigma_k \\ \Sigma_k^\dagger & b_k \end{pmatrix} \begin{pmatrix} W_k^\dagger & O \\ O & V_k^\dagger \end{pmatrix}
\end{equation}
where $b_k=V_k^{-1}B_k(V_k^\dagger)^{-1}$.
So that $H_k$ is similar to a matrix that contains $N_A - N_B$ zero rows, which implies that $H_k$ possesses at least $N_A - N_B$ zero-energy solutions with any $k$. Then we could conclude that a BCL has at least $N_A - N_B$ degenerate flat bands at the onsite energy of orbitals on $A$ sublattice. This proof does not make requirements on the form of $B_k$. Therefore, intra-sublattice hoppings and crystal field splitting of orbitals on the $B$ sublattice do not affect the flat band.

\bibliography{main}

\begin{thebibliography}{64}%
\makeatletter
\providecommand \@ifxundefined [1]{%
 \@ifx{#1\undefined}
}%
\providecommand \@ifnum [1]{%
 \ifnum #1\expandafter \@firstoftwo
 \else \expandafter \@secondoftwo
 \fi
}%
\providecommand \@ifx [1]{%
 \ifx #1\expandafter \@firstoftwo
 \else \expandafter \@secondoftwo
 \fi
}%
\providecommand \natexlab [1]{#1}%
\providecommand \enquote  [1]{``#1''}%
\providecommand \bibnamefont  [1]{#1}%
\providecommand \bibfnamefont [1]{#1}%
\providecommand \citenamefont [1]{#1}%
\providecommand \href@noop [0]{\@secondoftwo}%
\providecommand \href [0]{\begingroup \@sanitize@url \@href}%
\providecommand \@href[1]{\@@startlink{#1}\@@href}%
\providecommand \@@href[1]{\endgroup#1\@@endlink}%
\providecommand \@sanitize@url [0]{\catcode `\\12\catcode `\$12\catcode
  `\&12\catcode `\#12\catcode `\^12\catcode `\_12\catcode `\%12\relax}%
\providecommand \@@startlink[1]{}%
\providecommand \@@endlink[0]{}%
\providecommand \url  [0]{\begingroup\@sanitize@url \@url }%
\providecommand \@url [1]{\endgroup\@href {#1}{\urlprefix }}%
\providecommand \urlprefix  [0]{URL }%
\providecommand \Eprint [0]{\href }%
\providecommand \doibase [0]{https://doi.org/}%
\providecommand \selectlanguage [0]{\@gobble}%
\providecommand \bibinfo  [0]{\@secondoftwo}%
\providecommand \bibfield  [0]{\@secondoftwo}%
\providecommand \translation [1]{[#1]}%
\providecommand \BibitemOpen [0]{}%
\providecommand \bibitemStop [0]{}%
\providecommand \bibitemNoStop [0]{.\EOS\space}%
\providecommand \EOS [0]{\spacefactor3000\relax}%
\providecommand \BibitemShut  [1]{\csname bibitem#1\endcsname}%
\let\auto@bib@innerbib\@empty
\bibitem [{\citenamefont {Bera}\ \emph {et~al.}(2023)\citenamefont {Bera},
  \citenamefont {Pradhan}, \citenamefont {Khan}, \citenamefont {Pal},
  \citenamefont {Pal}, \citenamefont {Kalimuddin}, \citenamefont {Bera},
  \citenamefont {Das}, \citenamefont {Pal},\ and\ \citenamefont
  {Mondal}}]{Fe4m1}%
  \BibitemOpen
  \bibfield  {author} {\bibinfo {author} {\bibfnamefont {S.}~\bibnamefont
  {Bera}}, \bibinfo {author} {\bibfnamefont {S.~K.}\ \bibnamefont {Pradhan}},
  \bibinfo {author} {\bibfnamefont {M.~S.}\ \bibnamefont {Khan}}, \bibinfo
  {author} {\bibfnamefont {R.}~\bibnamefont {Pal}}, \bibinfo {author}
  {\bibfnamefont {B.}~\bibnamefont {Pal}}, \bibinfo {author} {\bibfnamefont
  {S.}~\bibnamefont {Kalimuddin}}, \bibinfo {author} {\bibfnamefont
  {A.}~\bibnamefont {Bera}}, \bibinfo {author} {\bibfnamefont {B.}~\bibnamefont
  {Das}}, \bibinfo {author} {\bibfnamefont {A.~N.}\ \bibnamefont {Pal}},\ and\
  \bibinfo {author} {\bibfnamefont {M.}~\bibnamefont {Mondal}},\ }\bibfield
  {title} {\bibinfo {title} {Unravelling the nature of spin reorientation
  transition in quasi-2d vdw magnetic material, {$\mathrm{Fe_4GeTe_2}$}},\
  }\href {https://doi.org/https://doi.org/10.1016/j.jmmm.2022.170257}
  {\bibfield  {journal} {\bibinfo  {journal} {Journal of Magnetism and Magnetic
  Materials}\ }\textbf {\bibinfo {volume} {565}},\ \bibinfo {pages} {170257}
  (\bibinfo {year} {2023})}\BibitemShut {NoStop}%
\bibitem [{\citenamefont {Chen}\ \emph
  {et~al.}(2022{\natexlab{a}})\citenamefont {Chen}, \citenamefont {Asif},
  \citenamefont {Whalen}, \citenamefont {Támara-Isaza}, \citenamefont
  {Luetke}, \citenamefont {Wang}, \citenamefont {Wang}, \citenamefont {Ayako},
  \citenamefont {Lamsal}, \citenamefont {May}, \citenamefont {McGuire},
  \citenamefont {Chakraborty}, \citenamefont {Xiao},\ and\ \citenamefont
  {Ku}}]{Fe5m1}%
  \BibitemOpen
  \bibfield  {author} {\bibinfo {author} {\bibfnamefont {H.}~\bibnamefont
  {Chen}}, \bibinfo {author} {\bibfnamefont {S.}~\bibnamefont {Asif}}, \bibinfo
  {author} {\bibfnamefont {M.}~\bibnamefont {Whalen}}, \bibinfo {author}
  {\bibfnamefont {J.}~\bibnamefont {Támara-Isaza}}, \bibinfo {author}
  {\bibfnamefont {B.}~\bibnamefont {Luetke}}, \bibinfo {author} {\bibfnamefont
  {Y.}~\bibnamefont {Wang}}, \bibinfo {author} {\bibfnamefont {X.}~\bibnamefont
  {Wang}}, \bibinfo {author} {\bibfnamefont {M.}~\bibnamefont {Ayako}},
  \bibinfo {author} {\bibfnamefont {S.}~\bibnamefont {Lamsal}}, \bibinfo
  {author} {\bibfnamefont {A.~F.}\ \bibnamefont {May}}, \bibinfo {author}
  {\bibfnamefont {M.~A.}\ \bibnamefont {McGuire}}, \bibinfo {author}
  {\bibfnamefont {C.}~\bibnamefont {Chakraborty}}, \bibinfo {author}
  {\bibfnamefont {J.~Q.}\ \bibnamefont {Xiao}},\ and\ \bibinfo {author}
  {\bibfnamefont {M.~J.~H.}\ \bibnamefont {Ku}},\ }\bibfield  {title} {\bibinfo
  {title} {Revealing room temperature ferromagnetism in exfoliated
  {$\mathrm{Fe_5GeTe_2}$} flakes with quantum magnetic imaging},\ }\href
  {https://doi.org/10.1088/2053-1583/ac57a9} {\bibfield  {journal} {\bibinfo
  {journal} {2D Materials}\ }\textbf {\bibinfo {volume} {9}},\ \bibinfo {pages}
  {025017} (\bibinfo {year} {2022}{\natexlab{a}})}\BibitemShut {NoStop}%
\bibitem [{\citenamefont {Chen}\ \emph {et~al.}(2023)\citenamefont {Chen},
  \citenamefont {Asif}, \citenamefont {Dolui}, \citenamefont {Wang},
  \citenamefont {Támara-Isaza}, \citenamefont {Goli}, \citenamefont {Whalen},
  \citenamefont {Wang}, \citenamefont {Chen}, \citenamefont {Zhang},
  \citenamefont {Liu}, \citenamefont {Jariwala}, \citenamefont {Jungfleisch},
  \citenamefont {Chakraborty}, \citenamefont {May}, \citenamefont {McGuire},
  \citenamefont {Nikolic}, \citenamefont {Xiao},\ and\ \citenamefont
  {Ku}}]{Fe5m2}%
  \BibitemOpen
  \bibfield  {author} {\bibinfo {author} {\bibfnamefont {H.}~\bibnamefont
  {Chen}}, \bibinfo {author} {\bibfnamefont {S.}~\bibnamefont {Asif}}, \bibinfo
  {author} {\bibfnamefont {K.}~\bibnamefont {Dolui}}, \bibinfo {author}
  {\bibfnamefont {Y.}~\bibnamefont {Wang}}, \bibinfo {author} {\bibfnamefont
  {J.}~\bibnamefont {Támara-Isaza}}, \bibinfo {author} {\bibfnamefont {V.~M.
  L. D.~P.}\ \bibnamefont {Goli}}, \bibinfo {author} {\bibfnamefont
  {M.}~\bibnamefont {Whalen}}, \bibinfo {author} {\bibfnamefont
  {X.}~\bibnamefont {Wang}}, \bibinfo {author} {\bibfnamefont {Z.}~\bibnamefont
  {Chen}}, \bibinfo {author} {\bibfnamefont {H.}~\bibnamefont {Zhang}},
  \bibinfo {author} {\bibfnamefont {K.}~\bibnamefont {Liu}}, \bibinfo {author}
  {\bibfnamefont {D.}~\bibnamefont {Jariwala}}, \bibinfo {author}
  {\bibfnamefont {M.~B.}\ \bibnamefont {Jungfleisch}}, \bibinfo {author}
  {\bibfnamefont {C.}~\bibnamefont {Chakraborty}}, \bibinfo {author}
  {\bibfnamefont {A.~F.}\ \bibnamefont {May}}, \bibinfo {author} {\bibfnamefont
  {M.~A.}\ \bibnamefont {McGuire}}, \bibinfo {author} {\bibfnamefont {B.~K.}\
  \bibnamefont {Nikolic}}, \bibinfo {author} {\bibfnamefont {J.~Q.}\
  \bibnamefont {Xiao}},\ and\ \bibinfo {author} {\bibfnamefont {M.~J.~H.}\
  \bibnamefont {Ku}},\ }\bibfield  {title} {\bibinfo {title}
  {Above-room-temperature ferromagnetism in thin van der waals flakes of
  cobalt-substituted {$\mathrm{Fe_5GeTe_2}$}},\ }\href
  {https://doi.org/10.1021/acsami.2c18028} {\bibfield  {journal} {\bibinfo
  {journal} {ACS Applied Materials}\ }\textbf {\bibinfo {volume} {15}},\
  \bibinfo {pages} {3287} (\bibinfo {year} {2023})}\BibitemShut {NoStop}%
\bibitem [{\citenamefont {Seo}\ \emph {et~al.}(2020)\citenamefont {Seo},
  \citenamefont {Kim}, \citenamefont {An}, \citenamefont {Kim}, \citenamefont
  {Kim}, \citenamefont {Hwang}, \citenamefont {Kim}, \citenamefont {Jang},
  \citenamefont {Kim}, \citenamefont {Eom}, \citenamefont {Seo}, \citenamefont
  {Stania}, \citenamefont {Muntwiler}, \citenamefont {Lee}, \citenamefont
  {Watanabe}, \citenamefont {Taniguchi}, \citenamefont {Jo}, \citenamefont
  {Lee}, \citenamefont {Min}, \citenamefont {Jo}, \citenamefont {Yeom},
  \citenamefont {Choi}, \citenamefont {Shim},\ and\ \citenamefont {Kim}}]{Fe4}%
  \BibitemOpen
  \bibfield  {author} {\bibinfo {author} {\bibfnamefont {J.}~\bibnamefont
  {Seo}}, \bibinfo {author} {\bibfnamefont {D.~Y.}\ \bibnamefont {Kim}},
  \bibinfo {author} {\bibfnamefont {E.~S.}\ \bibnamefont {An}}, \bibinfo
  {author} {\bibfnamefont {K.}~\bibnamefont {Kim}}, \bibinfo {author}
  {\bibfnamefont {G.~Y.}\ \bibnamefont {Kim}}, \bibinfo {author} {\bibfnamefont
  {S.~Y.}\ \bibnamefont {Hwang}}, \bibinfo {author} {\bibfnamefont {D.~W.}\
  \bibnamefont {Kim}}, \bibinfo {author} {\bibfnamefont {B.~G.}\ \bibnamefont
  {Jang}}, \bibinfo {author} {\bibfnamefont {H.}~\bibnamefont {Kim}}, \bibinfo
  {author} {\bibfnamefont {G.}~\bibnamefont {Eom}}, \bibinfo {author}
  {\bibfnamefont {S.~Y.}\ \bibnamefont {Seo}}, \bibinfo {author} {\bibfnamefont
  {R.}~\bibnamefont {Stania}}, \bibinfo {author} {\bibfnamefont
  {M.}~\bibnamefont {Muntwiler}}, \bibinfo {author} {\bibfnamefont
  {J.}~\bibnamefont {Lee}}, \bibinfo {author} {\bibfnamefont {K.}~\bibnamefont
  {Watanabe}}, \bibinfo {author} {\bibfnamefont {T.}~\bibnamefont {Taniguchi}},
  \bibinfo {author} {\bibfnamefont {Y.~J.}\ \bibnamefont {Jo}}, \bibinfo
  {author} {\bibfnamefont {J.}~\bibnamefont {Lee}}, \bibinfo {author}
  {\bibfnamefont {B.~I.}\ \bibnamefont {Min}}, \bibinfo {author} {\bibfnamefont
  {M.~H.}\ \bibnamefont {Jo}}, \bibinfo {author} {\bibfnamefont {H.~W.}\
  \bibnamefont {Yeom}}, \bibinfo {author} {\bibfnamefont {S.~Y.}\ \bibnamefont
  {Choi}}, \bibinfo {author} {\bibfnamefont {J.~H.}\ \bibnamefont {Shim}},\
  and\ \bibinfo {author} {\bibfnamefont {J.~S.}\ \bibnamefont {Kim}},\
  }\bibfield  {title} {\bibinfo {title} {Nearly room temperature ferromagnetism
  in a magnetic metal-rich van der waals metal},\ }\href
  {https://doi.org/10.1126/sciadv.aay8912} {\bibfield  {journal} {\bibinfo
  {journal} {Science Advances}\ }\textbf {\bibinfo {volume} {6}},\ \bibinfo
  {pages} {eaay8912} (\bibinfo {year} {2020})}\BibitemShut {NoStop}%
\bibitem [{\citenamefont {Stahl}\ \emph {et~al.}(2018)\citenamefont {Stahl},
  \citenamefont {Shlaen},\ and\ \citenamefont {Johrendt}}]{Fe5}%
  \BibitemOpen
  \bibfield  {author} {\bibinfo {author} {\bibfnamefont {J.}~\bibnamefont
  {Stahl}}, \bibinfo {author} {\bibfnamefont {E.}~\bibnamefont {Shlaen}},\ and\
  \bibinfo {author} {\bibfnamefont {D.}~\bibnamefont {Johrendt}},\ }\bibfield
  {title} {\bibinfo {title} {The van der waals ferromagnets
  {$\mathrm{Fe_{5-\delta}GeTe_2}$ and $\mathrm{Fe_{5-\delta -x}Ni_xGeTe_2}$}
  crystal structure, stacking faults, and magnetic properties},\ }\href
  {https://doi.org/10.1002/zaac.201800456} {\bibfield  {journal} {\bibinfo
  {journal} {Zeitschrift für anorganische und allgemeine Chemie}\ }\textbf
  {\bibinfo {volume} {644}},\ \bibinfo {pages} {1923} (\bibinfo {year}
  {2018})}\BibitemShut {NoStop}%
\bibitem [{\citenamefont {Wang}\ \emph {et~al.}(2023)\citenamefont {Wang},
  \citenamefont {Lu}, \citenamefont {Guo}, \citenamefont {Li}, \citenamefont
  {Wu}, \citenamefont {Li}, \citenamefont {Xie}, \citenamefont {Sun},
  \citenamefont {Li}, \citenamefont {Damas}, \citenamefont {Friedel},
  \citenamefont {Migot}, \citenamefont {Ghanbaja}, \citenamefont {Moreau},
  \citenamefont {Fagot-Revurat}, \citenamefont {Petit-Watelot}, \citenamefont
  {Hauet}, \citenamefont {Robertson}, \citenamefont {Mangin}, \citenamefont
  {Zhao},\ and\ \citenamefont {Nie}}]{Fe4n}%
  \BibitemOpen
  \bibfield  {author} {\bibinfo {author} {\bibfnamefont {H.}~\bibnamefont
  {Wang}}, \bibinfo {author} {\bibfnamefont {H.}~\bibnamefont {Lu}}, \bibinfo
  {author} {\bibfnamefont {Z.}~\bibnamefont {Guo}}, \bibinfo {author}
  {\bibfnamefont {A.}~\bibnamefont {Li}}, \bibinfo {author} {\bibfnamefont
  {P.}~\bibnamefont {Wu}}, \bibinfo {author} {\bibfnamefont {J.}~\bibnamefont
  {Li}}, \bibinfo {author} {\bibfnamefont {W.}~\bibnamefont {Xie}}, \bibinfo
  {author} {\bibfnamefont {Z.}~\bibnamefont {Sun}}, \bibinfo {author}
  {\bibfnamefont {P.}~\bibnamefont {Li}}, \bibinfo {author} {\bibfnamefont
  {H.}~\bibnamefont {Damas}}, \bibinfo {author} {\bibfnamefont {A.~M.}\
  \bibnamefont {Friedel}}, \bibinfo {author} {\bibfnamefont {S.}~\bibnamefont
  {Migot}}, \bibinfo {author} {\bibfnamefont {J.}~\bibnamefont {Ghanbaja}},
  \bibinfo {author} {\bibfnamefont {L.}~\bibnamefont {Moreau}}, \bibinfo
  {author} {\bibfnamefont {Y.}~\bibnamefont {Fagot-Revurat}}, \bibinfo {author}
  {\bibfnamefont {S.}~\bibnamefont {Petit-Watelot}}, \bibinfo {author}
  {\bibfnamefont {T.}~\bibnamefont {Hauet}}, \bibinfo {author} {\bibfnamefont
  {J.}~\bibnamefont {Robertson}}, \bibinfo {author} {\bibfnamefont
  {S.}~\bibnamefont {Mangin}}, \bibinfo {author} {\bibfnamefont
  {W.}~\bibnamefont {Zhao}},\ and\ \bibinfo {author} {\bibfnamefont
  {T.}~\bibnamefont {Nie}},\ }\bibfield  {title} {\bibinfo {title} {Interfacial
  engineering of ferromagnetism in wafer-scale van der waals
  {$\mathrm{Fe_4GeTe2}$} far above room temperature},\ }\href
  {https://doi.org/10.1038/s41467-023-37917-8} {\bibfield  {journal} {\bibinfo
  {journal} {Nature Communications}\ }\textbf {\bibinfo {volume} {14}},\
  \bibinfo {pages} {2483} (\bibinfo {year} {2023})}\BibitemShut {NoStop}%
\bibitem [{\citenamefont {Tang}\ \emph {et~al.}(2023)\citenamefont {Tang},
  \citenamefont {Huang}, \citenamefont {Qin}, \citenamefont {Zhai},
  \citenamefont {Ideue}, \citenamefont {Li}, \citenamefont {Meng},
  \citenamefont {Nie}, \citenamefont {Wu}, \citenamefont {Bi}, \citenamefont
  {Zhang}, \citenamefont {Zhou}, \citenamefont {Chen}, \citenamefont {Qiu},
  \citenamefont {Tang}, \citenamefont {Zhang}, \citenamefont {Wan},
  \citenamefont {Wang}, \citenamefont {Liu}, \citenamefont {Tian},
  \citenamefont {Iwasa},\ and\ \citenamefont {Yuan}}]{Fe5y}%
  \BibitemOpen
  \bibfield  {author} {\bibinfo {author} {\bibfnamefont {M.}~\bibnamefont
  {Tang}}, \bibinfo {author} {\bibfnamefont {J.}~\bibnamefont {Huang}},
  \bibinfo {author} {\bibfnamefont {F.}~\bibnamefont {Qin}}, \bibinfo {author}
  {\bibfnamefont {K.}~\bibnamefont {Zhai}}, \bibinfo {author} {\bibfnamefont
  {T.}~\bibnamefont {Ideue}}, \bibinfo {author} {\bibfnamefont
  {Z.}~\bibnamefont {Li}}, \bibinfo {author} {\bibfnamefont {F.}~\bibnamefont
  {Meng}}, \bibinfo {author} {\bibfnamefont {A.}~\bibnamefont {Nie}}, \bibinfo
  {author} {\bibfnamefont {L.}~\bibnamefont {Wu}}, \bibinfo {author}
  {\bibfnamefont {X.}~\bibnamefont {Bi}}, \bibinfo {author} {\bibfnamefont
  {C.}~\bibnamefont {Zhang}}, \bibinfo {author} {\bibfnamefont
  {L.}~\bibnamefont {Zhou}}, \bibinfo {author} {\bibfnamefont {P.}~\bibnamefont
  {Chen}}, \bibinfo {author} {\bibfnamefont {C.}~\bibnamefont {Qiu}}, \bibinfo
  {author} {\bibfnamefont {P.}~\bibnamefont {Tang}}, \bibinfo {author}
  {\bibfnamefont {H.}~\bibnamefont {Zhang}}, \bibinfo {author} {\bibfnamefont
  {X.}~\bibnamefont {Wan}}, \bibinfo {author} {\bibfnamefont {L.}~\bibnamefont
  {Wang}}, \bibinfo {author} {\bibfnamefont {Z.}~\bibnamefont {Liu}}, \bibinfo
  {author} {\bibfnamefont {Y.}~\bibnamefont {Tian}}, \bibinfo {author}
  {\bibfnamefont {Y.}~\bibnamefont {Iwasa}},\ and\ \bibinfo {author}
  {\bibfnamefont {H.}~\bibnamefont {Yuan}},\ }\bibfield  {title} {\bibinfo
  {title} {Continuous manipulation of magnetic anisotropy in a van der waals
  ferromagnet via electrical gating},\ }\href
  {https://doi.org/10.1038/s41928-022-00882-z} {\bibfield  {journal} {\bibinfo
  {journal} {Nature Electronics}\ }\textbf {\bibinfo {volume} {6}},\ \bibinfo
  {pages} {28} (\bibinfo {year} {2023})}\BibitemShut {NoStop}%
\bibitem [{\citenamefont {Seo}\ \emph {et~al.}(2021{\natexlab{a}})\citenamefont
  {Seo}, \citenamefont {An}, \citenamefont {Park}, \citenamefont {Hwang},
  \citenamefont {Kim}, \citenamefont {Song}, \citenamefont {Noh}, \citenamefont
  {Kim}, \citenamefont {Choi}, \citenamefont {Choi}, \citenamefont {Oh},
  \citenamefont {Watanabe}, \citenamefont {Taniguchi}, \citenamefont {Park},
  \citenamefont {Jo}, \citenamefont {Yeom}, \citenamefont {Choi}, \citenamefont
  {Shim},\ and\ \citenamefont {Kim}}]{Feaf}%
  \BibitemOpen
  \bibfield  {author} {\bibinfo {author} {\bibfnamefont {J.}~\bibnamefont
  {Seo}}, \bibinfo {author} {\bibfnamefont {E.~S.}\ \bibnamefont {An}},
  \bibinfo {author} {\bibfnamefont {T.}~\bibnamefont {Park}}, \bibinfo {author}
  {\bibfnamefont {S.-Y.}\ \bibnamefont {Hwang}}, \bibinfo {author}
  {\bibfnamefont {G.-Y.}\ \bibnamefont {Kim}}, \bibinfo {author} {\bibfnamefont
  {K.}~\bibnamefont {Song}}, \bibinfo {author} {\bibfnamefont {W.-s.}\
  \bibnamefont {Noh}}, \bibinfo {author} {\bibfnamefont {J.~Y.}\ \bibnamefont
  {Kim}}, \bibinfo {author} {\bibfnamefont {G.~S.}\ \bibnamefont {Choi}},
  \bibinfo {author} {\bibfnamefont {M.}~\bibnamefont {Choi}}, \bibinfo {author}
  {\bibfnamefont {E.}~\bibnamefont {Oh}}, \bibinfo {author} {\bibfnamefont
  {K.}~\bibnamefont {Watanabe}}, \bibinfo {author} {\bibfnamefont
  {T.}~\bibnamefont {Taniguchi}}, \bibinfo {author} {\bibfnamefont {J.~H.}\
  \bibnamefont {Park}}, \bibinfo {author} {\bibfnamefont {Y.~J.}\ \bibnamefont
  {Jo}}, \bibinfo {author} {\bibfnamefont {H.~W.}\ \bibnamefont {Yeom}},
  \bibinfo {author} {\bibfnamefont {S.-Y.}\ \bibnamefont {Choi}}, \bibinfo
  {author} {\bibfnamefont {J.~H.}\ \bibnamefont {Shim}},\ and\ \bibinfo
  {author} {\bibfnamefont {J.~S.}\ \bibnamefont {Kim}},\ }\bibfield  {title}
  {\bibinfo {title} {Tunable high-temperature itinerant antiferromagnetism in a
  van der waals magnet},\ }\href {https://doi.org/10.1038/s41467-021-23122-y}
  {\bibfield  {journal} {\bibinfo  {journal} {Nature Communications}\ }\textbf
  {\bibinfo {volume} {12}},\ \bibinfo {pages} {2844} (\bibinfo {year}
  {2021}{\natexlab{a}})}\BibitemShut {NoStop}%
\bibitem [{\citenamefont {May}\ \emph {et~al.}(2019{\natexlab{a}})\citenamefont
  {May}, \citenamefont {Bridges},\ and\ \citenamefont {McGuire}}]{Fe5pt}%
  \BibitemOpen
  \bibfield  {author} {\bibinfo {author} {\bibfnamefont {A.~F.}\ \bibnamefont
  {May}}, \bibinfo {author} {\bibfnamefont {C.~A.}\ \bibnamefont {Bridges}},\
  and\ \bibinfo {author} {\bibfnamefont {M.~A.}\ \bibnamefont {McGuire}},\
  }\bibfield  {title} {\bibinfo {title} {Physical properties and thermal
  stability of {${\mathrm{Fe}}_{5\ensuremath{-}x}{\mathrm{GeTe}}_{2}$} single
  crystals},\ }\href {https://doi.org/10.1103/PhysRevMaterials.3.104401}
  {\bibfield  {journal} {\bibinfo  {journal} {Physical Review Materials}\
  }\textbf {\bibinfo {volume} {3}},\ \bibinfo {pages} {104401} (\bibinfo {year}
  {2019}{\natexlab{a}})}\BibitemShut {NoStop}%
\bibitem [{\citenamefont {Chen}\ \emph
  {et~al.}(2022{\natexlab{b}})\citenamefont {Chen}, \citenamefont {Shao},
  \citenamefont {Chen}, \citenamefont {Susarla}, \citenamefont {Hogan},
  \citenamefont {He}, \citenamefont {Zhang}, \citenamefont {Wang},
  \citenamefont {Yao}, \citenamefont {Ercius}, \citenamefont {Muller},
  \citenamefont {Ramesh},\ and\ \citenamefont {Birgeneau}}]{Fe2dm}%
  \BibitemOpen
  \bibfield  {author} {\bibinfo {author} {\bibfnamefont {X.}~\bibnamefont
  {Chen}}, \bibinfo {author} {\bibfnamefont {Y.-T.}\ \bibnamefont {Shao}},
  \bibinfo {author} {\bibfnamefont {R.}~\bibnamefont {Chen}}, \bibinfo {author}
  {\bibfnamefont {S.}~\bibnamefont {Susarla}}, \bibinfo {author} {\bibfnamefont
  {T.}~\bibnamefont {Hogan}}, \bibinfo {author} {\bibfnamefont
  {Y.}~\bibnamefont {He}}, \bibinfo {author} {\bibfnamefont {H.}~\bibnamefont
  {Zhang}}, \bibinfo {author} {\bibfnamefont {S.}~\bibnamefont {Wang}},
  \bibinfo {author} {\bibfnamefont {J.}~\bibnamefont {Yao}}, \bibinfo {author}
  {\bibfnamefont {P.}~\bibnamefont {Ercius}}, \bibinfo {author} {\bibfnamefont
  {D.~A.}\ \bibnamefont {Muller}}, \bibinfo {author} {\bibfnamefont
  {R.}~\bibnamefont {Ramesh}},\ and\ \bibinfo {author} {\bibfnamefont {R.~J.}\
  \bibnamefont {Birgeneau}},\ }\bibfield  {title} {\bibinfo {title} {Pervasive
  beyond room-temperature ferromagnetism in a doped van der waals magnet},\
  }\href {https://doi.org/10.1103/PhysRevLett.128.217203} {\bibfield  {journal}
  {\bibinfo  {journal} {Physical Review Letters}\ }\textbf {\bibinfo {volume}
  {128}},\ \bibinfo {pages} {217203} (\bibinfo {year}
  {2022}{\natexlab{b}})}\BibitemShut {NoStop}%
\bibitem [{\citenamefont {Ghosh}\ \emph {et~al.}(2023)\citenamefont {Ghosh},
  \citenamefont {Ershadrad}, \citenamefont {Borisov},\ and\ \citenamefont
  {Sanyal}}]{Fel345}%
  \BibitemOpen
  \bibfield  {author} {\bibinfo {author} {\bibfnamefont {S.}~\bibnamefont
  {Ghosh}}, \bibinfo {author} {\bibfnamefont {S.}~\bibnamefont {Ershadrad}},
  \bibinfo {author} {\bibfnamefont {V.}~\bibnamefont {Borisov}},\ and\ \bibinfo
  {author} {\bibfnamefont {B.}~\bibnamefont {Sanyal}},\ }\bibfield  {title}
  {\bibinfo {title} {Unraveling effects of electron correlation in
  two-dimensional {$\mathrm{Fe_nGeTe_2 (n = 3, 4, 5)}$} by dynamical mean field
  theory},\ }\href {https://doi.org/10.1038/s41524-023-01024-5} {\bibfield
  {journal} {\bibinfo  {journal} {npj Computational Materials}\ }\textbf
  {\bibinfo {volume} {9}},\ \bibinfo {pages} {86} (\bibinfo {year}
  {2023})}\BibitemShut {NoStop}%
\bibitem [{\citenamefont {Mondal}\ \emph {et~al.}(2021)\citenamefont {Mondal},
  \citenamefont {Khan}, \citenamefont {Mishra}, \citenamefont {Satpati},\ and\
  \citenamefont {Mandal}}]{Fe42}%
  \BibitemOpen
  \bibfield  {author} {\bibinfo {author} {\bibfnamefont {S.}~\bibnamefont
  {Mondal}}, \bibinfo {author} {\bibfnamefont {N.}~\bibnamefont {Khan}},
  \bibinfo {author} {\bibfnamefont {S.~M.}\ \bibnamefont {Mishra}}, \bibinfo
  {author} {\bibfnamefont {B.}~\bibnamefont {Satpati}},\ and\ \bibinfo {author}
  {\bibfnamefont {P.}~\bibnamefont {Mandal}},\ }\bibfield  {title} {\bibinfo
  {title} {Critical behavior in the van der waals itinerant ferromagnet
  {${\mathrm{Fe}}_{4}{\mathrm{GeTe}}_{2}$}},\ }\href
  {https://doi.org/10.1103/PhysRevB.104.094405} {\bibfield  {journal} {\bibinfo
   {journal} {Physical Review B}\ }\textbf {\bibinfo {volume} {104}},\ \bibinfo
  {pages} {094405} (\bibinfo {year} {2021})}\BibitemShut {NoStop}%
\bibitem [{\citenamefont {May}\ \emph {et~al.}(2019{\natexlab{b}})\citenamefont
  {May}, \citenamefont {Ovchinnikov}, \citenamefont {Zheng}, \citenamefont
  {Hermann}, \citenamefont {Calder}, \citenamefont {Huang}, \citenamefont
  {Fei}, \citenamefont {Liu}, \citenamefont {Xu},\ and\ \citenamefont
  {McGuire}}]{Fe52}%
  \BibitemOpen
  \bibfield  {author} {\bibinfo {author} {\bibfnamefont {A.~F.}\ \bibnamefont
  {May}}, \bibinfo {author} {\bibfnamefont {D.}~\bibnamefont {Ovchinnikov}},
  \bibinfo {author} {\bibfnamefont {Q.}~\bibnamefont {Zheng}}, \bibinfo
  {author} {\bibfnamefont {R.}~\bibnamefont {Hermann}}, \bibinfo {author}
  {\bibfnamefont {S.}~\bibnamefont {Calder}}, \bibinfo {author} {\bibfnamefont
  {B.}~\bibnamefont {Huang}}, \bibinfo {author} {\bibfnamefont
  {Z.}~\bibnamefont {Fei}}, \bibinfo {author} {\bibfnamefont {Y.}~\bibnamefont
  {Liu}}, \bibinfo {author} {\bibfnamefont {X.}~\bibnamefont {Xu}},\ and\
  \bibinfo {author} {\bibfnamefont {M.~A.}\ \bibnamefont {McGuire}},\
  }\bibfield  {title} {\bibinfo {title} {Ferromagnetism near room temperature
  in the cleavable van der waals crystal {$\mathrm{Fe_5GeTe_2}$}},\ }\href
  {https://doi.org/10.1021/acsnano.8b09660} {\bibfield  {journal} {\bibinfo
  {journal} {ACS Nano}\ }\textbf {\bibinfo {volume} {13}},\ \bibinfo {pages}
  {4436} (\bibinfo {year} {2019}{\natexlab{b}})}\BibitemShut {NoStop}%
\bibitem [{\citenamefont {Alahmed}\ \emph {et~al.}(2021)\citenamefont
  {Alahmed}, \citenamefont {Nepal}, \citenamefont {Macy}, \citenamefont
  {Zheng}, \citenamefont {Casas}, \citenamefont {Sapkota}, \citenamefont
  {Jones}, \citenamefont {Mazza}, \citenamefont {Brahlek}, \citenamefont {Jin},
  \citenamefont {Mahjouri-Samani}, \citenamefont {Zhang}, \citenamefont
  {Mewes}, \citenamefont {Balicas}, \citenamefont {Mewes},\ and\ \citenamefont
  {Li}}]{Fe53}%
  \BibitemOpen
  \bibfield  {author} {\bibinfo {author} {\bibfnamefont {L.}~\bibnamefont
  {Alahmed}}, \bibinfo {author} {\bibfnamefont {B.}~\bibnamefont {Nepal}},
  \bibinfo {author} {\bibfnamefont {J.}~\bibnamefont {Macy}}, \bibinfo {author}
  {\bibfnamefont {W.}~\bibnamefont {Zheng}}, \bibinfo {author} {\bibfnamefont
  {B.}~\bibnamefont {Casas}}, \bibinfo {author} {\bibfnamefont
  {A.}~\bibnamefont {Sapkota}}, \bibinfo {author} {\bibfnamefont
  {N.}~\bibnamefont {Jones}}, \bibinfo {author} {\bibfnamefont {A.~R.}\
  \bibnamefont {Mazza}}, \bibinfo {author} {\bibfnamefont {M.}~\bibnamefont
  {Brahlek}}, \bibinfo {author} {\bibfnamefont {W.}~\bibnamefont {Jin}},
  \bibinfo {author} {\bibfnamefont {M.}~\bibnamefont {Mahjouri-Samani}},
  \bibinfo {author} {\bibfnamefont {S.~S.~L.}\ \bibnamefont {Zhang}}, \bibinfo
  {author} {\bibfnamefont {C.}~\bibnamefont {Mewes}}, \bibinfo {author}
  {\bibfnamefont {L.}~\bibnamefont {Balicas}}, \bibinfo {author} {\bibfnamefont
  {T.}~\bibnamefont {Mewes}},\ and\ \bibinfo {author} {\bibfnamefont
  {P.}~\bibnamefont {Li}},\ }\bibfield  {title} {\bibinfo {title} {Magnetism
  and spin dynamics in room-temperature van der waals magnet
  {$\mathrm{Fe_5GeTe_2}$}},\ }\href {https://doi.org/10.1088/2053-1583/ac2028}
  {\bibfield  {journal} {\bibinfo  {journal} {2D Materials}\ }\textbf {\bibinfo
  {volume} {8}},\ \bibinfo {pages} {045030} (\bibinfo {year}
  {2021})}\BibitemShut {NoStop}%
\bibitem [{\citenamefont {Zhang}\ \emph {et~al.}(2020)\citenamefont {Zhang},
  \citenamefont {Chen}, \citenamefont {Zhai}, \citenamefont {Chen},
  \citenamefont {Caretta}, \citenamefont {Huang}, \citenamefont {Chopdekar},
  \citenamefont {Cao}, \citenamefont {Sun}, \citenamefont {Yao}, \citenamefont
  {Birgeneau},\ and\ \citenamefont {Ramesh}}]{Fe5m5}%
  \BibitemOpen
  \bibfield  {author} {\bibinfo {author} {\bibfnamefont {H.}~\bibnamefont
  {Zhang}}, \bibinfo {author} {\bibfnamefont {R.}~\bibnamefont {Chen}},
  \bibinfo {author} {\bibfnamefont {K.}~\bibnamefont {Zhai}}, \bibinfo {author}
  {\bibfnamefont {X.}~\bibnamefont {Chen}}, \bibinfo {author} {\bibfnamefont
  {L.}~\bibnamefont {Caretta}}, \bibinfo {author} {\bibfnamefont
  {X.}~\bibnamefont {Huang}}, \bibinfo {author} {\bibfnamefont {R.~V.}\
  \bibnamefont {Chopdekar}}, \bibinfo {author} {\bibfnamefont {J.}~\bibnamefont
  {Cao}}, \bibinfo {author} {\bibfnamefont {J.}~\bibnamefont {Sun}}, \bibinfo
  {author} {\bibfnamefont {J.}~\bibnamefont {Yao}}, \bibinfo {author}
  {\bibfnamefont {R.}~\bibnamefont {Birgeneau}},\ and\ \bibinfo {author}
  {\bibfnamefont {R.}~\bibnamefont {Ramesh}},\ }\bibfield  {title} {\bibinfo
  {title} {Itinerant ferromagnetism in van der waals
  {${\mathrm{Fe}}_{5\ensuremath{-}x}{\mathrm{Ge}\mathrm{Te}}_{2}$} crystals
  above room temperature},\ }\href
  {https://doi.org/10.1103/PhysRevB.102.064417} {\bibfield  {journal} {\bibinfo
   {journal} {Physical Review B}\ }\textbf {\bibinfo {volume} {102}},\ \bibinfo
  {pages} {064417} (\bibinfo {year} {2020})}\BibitemShut {NoStop}%
\bibitem [{\citenamefont {Liu}\ \emph {et~al.}(2022)\citenamefont {Liu},
  \citenamefont {Xing}, \citenamefont {Jiang}, \citenamefont {Guo},
  \citenamefont {Jiang}, \citenamefont {Qi},\ and\ \citenamefont
  {Zhao}}]{Felm}%
  \BibitemOpen
  \bibfield  {author} {\bibinfo {author} {\bibfnamefont {Q.}~\bibnamefont
  {Liu}}, \bibinfo {author} {\bibfnamefont {J.}~\bibnamefont {Xing}}, \bibinfo
  {author} {\bibfnamefont {Z.}~\bibnamefont {Jiang}}, \bibinfo {author}
  {\bibfnamefont {Y.}~\bibnamefont {Guo}}, \bibinfo {author} {\bibfnamefont
  {X.}~\bibnamefont {Jiang}}, \bibinfo {author} {\bibfnamefont
  {Y.}~\bibnamefont {Qi}},\ and\ \bibinfo {author} {\bibfnamefont
  {J.}~\bibnamefont {Zhao}},\ }\bibfield  {title} {\bibinfo {title}
  {Layer-dependent magnetic phase diagram in $\mathrm{Fe_nGeTe_2 (3 \le n \le
  7)}$ ultrathin films},\ }\href {https://doi.org/10.1038/s42005-022-00921-3}
  {\bibfield  {journal} {\bibinfo  {journal} {Communications Physics}\ }\textbf
  {\bibinfo {volume} {5}},\ \bibinfo {pages} {140} (\bibinfo {year}
  {2022})}\BibitemShut {NoStop}%
\bibitem [{\citenamefont {Tan}\ \emph {et~al.}(2021)\citenamefont {Tan},
  \citenamefont {Xie}, \citenamefont {Zheng}, \citenamefont {Aloufi},
  \citenamefont {Albarakati}, \citenamefont {Algarni}, \citenamefont {Li},
  \citenamefont {Partridge}, \citenamefont {Culcer}, \citenamefont {Wang},
  \citenamefont {Yi}, \citenamefont {Tian}, \citenamefont {Xiong},
  \citenamefont {Zhao},\ and\ \citenamefont {Wang}}]{Fegm}%
  \BibitemOpen
  \bibfield  {author} {\bibinfo {author} {\bibfnamefont {C.}~\bibnamefont
  {Tan}}, \bibinfo {author} {\bibfnamefont {W.-Q.}\ \bibnamefont {Xie}},
  \bibinfo {author} {\bibfnamefont {G.}~\bibnamefont {Zheng}}, \bibinfo
  {author} {\bibfnamefont {N.}~\bibnamefont {Aloufi}}, \bibinfo {author}
  {\bibfnamefont {S.}~\bibnamefont {Albarakati}}, \bibinfo {author}
  {\bibfnamefont {M.}~\bibnamefont {Algarni}}, \bibinfo {author} {\bibfnamefont
  {J.}~\bibnamefont {Li}}, \bibinfo {author} {\bibfnamefont {J.}~\bibnamefont
  {Partridge}}, \bibinfo {author} {\bibfnamefont {D.}~\bibnamefont {Culcer}},
  \bibinfo {author} {\bibfnamefont {X.}~\bibnamefont {Wang}}, \bibinfo {author}
  {\bibfnamefont {J.~B.}\ \bibnamefont {Yi}}, \bibinfo {author} {\bibfnamefont
  {M.}~\bibnamefont {Tian}}, \bibinfo {author} {\bibfnamefont {Y.}~\bibnamefont
  {Xiong}}, \bibinfo {author} {\bibfnamefont {Y.-J.}\ \bibnamefont {Zhao}},\
  and\ \bibinfo {author} {\bibfnamefont {L.}~\bibnamefont {Wang}},\ }\bibfield
  {title} {\bibinfo {title} {Gate-controlled magnetic phase transition in a van
  der waals magnet {$\mathrm{Fe_5GeTe_2}$}},\ }\href
  {https://doi.org/10.1021/acs.nanolett.1c01108} {\bibfield  {journal}
  {\bibinfo  {journal} {Nano Letters}\ }\textbf {\bibinfo {volume} {21}},\
  \bibinfo {pages} {5599} (\bibinfo {year} {2021})}\BibitemShut {NoStop}%
\bibitem [{\citenamefont {Seo}\ \emph {et~al.}(2021{\natexlab{b}})\citenamefont
  {Seo}, \citenamefont {An}, \citenamefont {Park}, \citenamefont {Hwang},
  \citenamefont {Kim}, \citenamefont {Song}, \citenamefont {Noh}, \citenamefont
  {Kim}, \citenamefont {Choi}, \citenamefont {Choi}, \citenamefont {Oh},
  \citenamefont {Watanabe}, \citenamefont {Taniguchi}, \citenamefont {Park},
  \citenamefont {Jo}, \citenamefont {Yeom}, \citenamefont {Choi}, \citenamefont
  {Shim},\ and\ \citenamefont {Kim}}]{Fe2mm}%
  \BibitemOpen
  \bibfield  {author} {\bibinfo {author} {\bibfnamefont {J.}~\bibnamefont
  {Seo}}, \bibinfo {author} {\bibfnamefont {E.~S.}\ \bibnamefont {An}},
  \bibinfo {author} {\bibfnamefont {T.}~\bibnamefont {Park}}, \bibinfo {author}
  {\bibfnamefont {S.-Y.}\ \bibnamefont {Hwang}}, \bibinfo {author}
  {\bibfnamefont {G.-Y.}\ \bibnamefont {Kim}}, \bibinfo {author} {\bibfnamefont
  {K.}~\bibnamefont {Song}}, \bibinfo {author} {\bibfnamefont {W.-s.}\
  \bibnamefont {Noh}}, \bibinfo {author} {\bibfnamefont {J.~Y.}\ \bibnamefont
  {Kim}}, \bibinfo {author} {\bibfnamefont {G.~S.}\ \bibnamefont {Choi}},
  \bibinfo {author} {\bibfnamefont {M.}~\bibnamefont {Choi}}, \bibinfo {author}
  {\bibfnamefont {E.}~\bibnamefont {Oh}}, \bibinfo {author} {\bibfnamefont
  {K.}~\bibnamefont {Watanabe}}, \bibinfo {author} {\bibfnamefont
  {T.}~\bibnamefont {Taniguchi}}, \bibinfo {author} {\bibfnamefont {J.~H.}\
  \bibnamefont {Park}}, \bibinfo {author} {\bibfnamefont {Y.~J.}\ \bibnamefont
  {Jo}}, \bibinfo {author} {\bibfnamefont {H.~W.}\ \bibnamefont {Yeom}},
  \bibinfo {author} {\bibfnamefont {S.-Y.}\ \bibnamefont {Choi}}, \bibinfo
  {author} {\bibfnamefont {J.~H.}\ \bibnamefont {Shim}},\ and\ \bibinfo
  {author} {\bibfnamefont {J.~S.}\ \bibnamefont {Kim}},\ }\bibfield  {title}
  {\bibinfo {title} {Tunable high-temperature itinerant antiferromagnetism in a
  van der waals magnet},\ }\href {https://doi.org/10.1038/s41467-021-23122-y}
  {\bibfield  {journal} {\bibinfo  {journal} {Nature Communications}\ }\textbf
  {\bibinfo {volume} {12}},\ \bibinfo {pages} {2844} (\bibinfo {year}
  {2021}{\natexlab{b}})}\BibitemShut {NoStop}%
\bibitem [{\citenamefont {Gong}\ and\ \citenamefont {Zhang}(2019)}]{2Dm1}%
  \BibitemOpen
  \bibfield  {author} {\bibinfo {author} {\bibfnamefont {C.}~\bibnamefont
  {Gong}}\ and\ \bibinfo {author} {\bibfnamefont {X.}~\bibnamefont {Zhang}},\
  }\bibfield  {title} {\bibinfo {title} {Two-dimensional magnetic crystals and
  emergent heterostructure devices},\ }\href
  {https://doi.org/10.1126/science.aav4450} {\bibfield  {journal} {\bibinfo
  {journal} {Science}\ }\textbf {\bibinfo {volume} {363}},\ \bibinfo {pages}
  {eaav4450} (\bibinfo {year} {2019})}\BibitemShut {NoStop}%
\bibitem [{\citenamefont {Burch}\ \emph {et~al.}(2018)\citenamefont {Burch},
  \citenamefont {Mandrus},\ and\ \citenamefont {Park}}]{2Dm2}%
  \BibitemOpen
  \bibfield  {author} {\bibinfo {author} {\bibfnamefont {K.~S.}\ \bibnamefont
  {Burch}}, \bibinfo {author} {\bibfnamefont {D.}~\bibnamefont {Mandrus}},\
  and\ \bibinfo {author} {\bibfnamefont {J.}~\bibnamefont {Park}},\ }\bibfield
  {title} {\bibinfo {title} {Magnetism in two-dimensional van der waals
  materials},\ }\href {https://doi.org/10.1038/s41586-018-0631-z} {\bibfield
  {journal} {\bibinfo  {journal} {Nature}\ }\textbf {\bibinfo {volume} {563}},\
  \bibinfo {pages} {47} (\bibinfo {year} {2018})}\BibitemShut {NoStop}%
\bibitem [{\citenamefont {Gong}\ \emph {et~al.}(2017)\citenamefont {Gong},
  \citenamefont {Li}, \citenamefont {Li}, \citenamefont {Ji}, \citenamefont
  {Stern}, \citenamefont {Xia}, \citenamefont {Cao}, \citenamefont {Bao},
  \citenamefont {Wang}, \citenamefont {Wang}, \citenamefont {Qiu},
  \citenamefont {Cava}, \citenamefont {Louie}, \citenamefont {Xia},\ and\
  \citenamefont {Zhang}}]{2Dm3}%
  \BibitemOpen
  \bibfield  {author} {\bibinfo {author} {\bibfnamefont {C.}~\bibnamefont
  {Gong}}, \bibinfo {author} {\bibfnamefont {L.}~\bibnamefont {Li}}, \bibinfo
  {author} {\bibfnamefont {Z.}~\bibnamefont {Li}}, \bibinfo {author}
  {\bibfnamefont {H.}~\bibnamefont {Ji}}, \bibinfo {author} {\bibfnamefont
  {A.}~\bibnamefont {Stern}}, \bibinfo {author} {\bibfnamefont
  {Y.}~\bibnamefont {Xia}}, \bibinfo {author} {\bibfnamefont {T.}~\bibnamefont
  {Cao}}, \bibinfo {author} {\bibfnamefont {W.}~\bibnamefont {Bao}}, \bibinfo
  {author} {\bibfnamefont {C.}~\bibnamefont {Wang}}, \bibinfo {author}
  {\bibfnamefont {Y.}~\bibnamefont {Wang}}, \bibinfo {author} {\bibfnamefont
  {Z.~Q.}\ \bibnamefont {Qiu}}, \bibinfo {author} {\bibfnamefont {R.~J.}\
  \bibnamefont {Cava}}, \bibinfo {author} {\bibfnamefont {S.~G.}\ \bibnamefont
  {Louie}}, \bibinfo {author} {\bibfnamefont {J.}~\bibnamefont {Xia}},\ and\
  \bibinfo {author} {\bibfnamefont {X.}~\bibnamefont {Zhang}},\ }\bibfield
  {title} {\bibinfo {title} {Discovery of intrinsic ferromagnetism in
  two-dimensional van der waals crystals},\ }\href
  {https://doi.org/10.1038/nature22060} {\bibfield  {journal} {\bibinfo
  {journal} {Nature}\ }\textbf {\bibinfo {volume} {546}},\ \bibinfo {pages}
  {265} (\bibinfo {year} {2017})}\BibitemShut {NoStop}%
\bibitem [{\citenamefont {Huang}\ \emph {et~al.}(2017)\citenamefont {Huang},
  \citenamefont {Clark}, \citenamefont {Navarro-Moratalla}, \citenamefont
  {Klein}, \citenamefont {Cheng}, \citenamefont {Seyler}, \citenamefont
  {Zhong}, \citenamefont {Schmidgall}, \citenamefont {McGuire}, \citenamefont
  {Cobden}, \citenamefont {Yao}, \citenamefont {Xiao}, \citenamefont
  {Jarillo-Herrero},\ and\ \citenamefont {Xu}}]{2Dm4}%
  \BibitemOpen
  \bibfield  {author} {\bibinfo {author} {\bibfnamefont {B.}~\bibnamefont
  {Huang}}, \bibinfo {author} {\bibfnamefont {G.}~\bibnamefont {Clark}},
  \bibinfo {author} {\bibfnamefont {E.}~\bibnamefont {Navarro-Moratalla}},
  \bibinfo {author} {\bibfnamefont {D.~R.}\ \bibnamefont {Klein}}, \bibinfo
  {author} {\bibfnamefont {R.}~\bibnamefont {Cheng}}, \bibinfo {author}
  {\bibfnamefont {K.~L.}\ \bibnamefont {Seyler}}, \bibinfo {author}
  {\bibfnamefont {D.}~\bibnamefont {Zhong}}, \bibinfo {author} {\bibfnamefont
  {E.}~\bibnamefont {Schmidgall}}, \bibinfo {author} {\bibfnamefont {M.~A.}\
  \bibnamefont {McGuire}}, \bibinfo {author} {\bibfnamefont {D.~H.}\
  \bibnamefont {Cobden}}, \bibinfo {author} {\bibfnamefont {W.}~\bibnamefont
  {Yao}}, \bibinfo {author} {\bibfnamefont {D.}~\bibnamefont {Xiao}}, \bibinfo
  {author} {\bibfnamefont {P.}~\bibnamefont {Jarillo-Herrero}},\ and\ \bibinfo
  {author} {\bibfnamefont {X.}~\bibnamefont {Xu}},\ }\bibfield  {title}
  {\bibinfo {title} {Layer-dependent ferromagnetism in a van der waals crystal
  down to the monolayer limit},\ }\href {https://doi.org/10.1038/nature22391}
  {\bibfield  {journal} {\bibinfo  {journal} {Nature}\ }\textbf {\bibinfo
  {volume} {546}},\ \bibinfo {pages} {270} (\bibinfo {year}
  {2017})}\BibitemShut {NoStop}%
\bibitem [{\citenamefont {Chen}\ \emph
  {et~al.}(2022{\natexlab{c}})\citenamefont {Chen}, \citenamefont {Shao},
  \citenamefont {Chen}, \citenamefont {Susarla}, \citenamefont {Hogan},
  \citenamefont {He}, \citenamefont {Zhang}, \citenamefont {Wang},
  \citenamefont {Yao}, \citenamefont {Ercius}, \citenamefont {Muller},
  \citenamefont {Ramesh},\ and\ \citenamefont {Birgeneau}}]{2Dm5}%
  \BibitemOpen
  \bibfield  {author} {\bibinfo {author} {\bibfnamefont {X.}~\bibnamefont
  {Chen}}, \bibinfo {author} {\bibfnamefont {Y.-T.}\ \bibnamefont {Shao}},
  \bibinfo {author} {\bibfnamefont {R.}~\bibnamefont {Chen}}, \bibinfo {author}
  {\bibfnamefont {S.}~\bibnamefont {Susarla}}, \bibinfo {author} {\bibfnamefont
  {T.}~\bibnamefont {Hogan}}, \bibinfo {author} {\bibfnamefont
  {Y.}~\bibnamefont {He}}, \bibinfo {author} {\bibfnamefont {H.}~\bibnamefont
  {Zhang}}, \bibinfo {author} {\bibfnamefont {S.}~\bibnamefont {Wang}},
  \bibinfo {author} {\bibfnamefont {J.}~\bibnamefont {Yao}}, \bibinfo {author}
  {\bibfnamefont {P.}~\bibnamefont {Ercius}}, \bibinfo {author} {\bibfnamefont
  {D.~A.}\ \bibnamefont {Muller}}, \bibinfo {author} {\bibfnamefont
  {R.}~\bibnamefont {Ramesh}},\ and\ \bibinfo {author} {\bibfnamefont {R.~J.}\
  \bibnamefont {Birgeneau}},\ }\bibfield  {title} {\bibinfo {title} {Pervasive
  beyond room-temperature ferromagnetism in a doped van der waals magnet},\
  }\href {https://doi.org/10.1103/PhysRevLett.128.217203} {\bibfield  {journal}
  {\bibinfo  {journal} {Physical Review Letters}\ }\textbf {\bibinfo {volume}
  {128}},\ \bibinfo {pages} {217203} (\bibinfo {year}
  {2022}{\natexlab{c}})}\BibitemShut {NoStop}%
\bibitem [{\citenamefont {Deng}\ \emph {et~al.}(2018)\citenamefont {Deng},
  \citenamefont {Yu}, \citenamefont {Song}, \citenamefont {Zhang},
  \citenamefont {Wang}, \citenamefont {Sun}, \citenamefont {Yi}, \citenamefont
  {Wu}, \citenamefont {Wu}, \citenamefont {Zhu}, \citenamefont {Wang},
  \citenamefont {Chen},\ and\ \citenamefont {Zhang}}]{Fe3N}%
  \BibitemOpen
  \bibfield  {author} {\bibinfo {author} {\bibfnamefont {Y.}~\bibnamefont
  {Deng}}, \bibinfo {author} {\bibfnamefont {Y.}~\bibnamefont {Yu}}, \bibinfo
  {author} {\bibfnamefont {Y.}~\bibnamefont {Song}}, \bibinfo {author}
  {\bibfnamefont {J.}~\bibnamefont {Zhang}}, \bibinfo {author} {\bibfnamefont
  {N.~Z.}\ \bibnamefont {Wang}}, \bibinfo {author} {\bibfnamefont
  {Z.}~\bibnamefont {Sun}}, \bibinfo {author} {\bibfnamefont {Y.}~\bibnamefont
  {Yi}}, \bibinfo {author} {\bibfnamefont {Y.~Z.}\ \bibnamefont {Wu}}, \bibinfo
  {author} {\bibfnamefont {S.}~\bibnamefont {Wu}}, \bibinfo {author}
  {\bibfnamefont {J.}~\bibnamefont {Zhu}}, \bibinfo {author} {\bibfnamefont
  {J.}~\bibnamefont {Wang}}, \bibinfo {author} {\bibfnamefont {X.~H.}\
  \bibnamefont {Chen}},\ and\ \bibinfo {author} {\bibfnamefont
  {Y.}~\bibnamefont {Zhang}},\ }\bibfield  {title} {\bibinfo {title}
  {Gate-tunable room-temperature ferromagnetism in two-dimensional
  {$\mathrm{Fe_3GeTe_2}$}},\ }\href {https://doi.org/10.1038/s41586-018-0626-9}
  {\bibfield  {journal} {\bibinfo  {journal} {Nature}\ }\textbf {\bibinfo
  {volume} {563}},\ \bibinfo {pages} {94} (\bibinfo {year} {2018})}\BibitemShut
  {NoStop}%
\bibitem [{\citenamefont {Fei}\ \emph {et~al.}(2018)\citenamefont {Fei},
  \citenamefont {Huang}, \citenamefont {Malinowski}, \citenamefont {Wang},
  \citenamefont {Song}, \citenamefont {Sanchez}, \citenamefont {Yao},
  \citenamefont {Xiao}, \citenamefont {Zhu}, \citenamefont {May}, \citenamefont
  {Wu}, \citenamefont {Cobden}, \citenamefont {Chu},\ and\ \citenamefont
  {Xu}}]{Fe3NM}%
  \BibitemOpen
  \bibfield  {author} {\bibinfo {author} {\bibfnamefont {Z.}~\bibnamefont
  {Fei}}, \bibinfo {author} {\bibfnamefont {B.}~\bibnamefont {Huang}}, \bibinfo
  {author} {\bibfnamefont {P.}~\bibnamefont {Malinowski}}, \bibinfo {author}
  {\bibfnamefont {W.}~\bibnamefont {Wang}}, \bibinfo {author} {\bibfnamefont
  {T.}~\bibnamefont {Song}}, \bibinfo {author} {\bibfnamefont {J.}~\bibnamefont
  {Sanchez}}, \bibinfo {author} {\bibfnamefont {W.}~\bibnamefont {Yao}},
  \bibinfo {author} {\bibfnamefont {D.}~\bibnamefont {Xiao}}, \bibinfo {author}
  {\bibfnamefont {X.}~\bibnamefont {Zhu}}, \bibinfo {author} {\bibfnamefont
  {A.~F.}\ \bibnamefont {May}}, \bibinfo {author} {\bibfnamefont
  {W.}~\bibnamefont {Wu}}, \bibinfo {author} {\bibfnamefont {D.~H.}\
  \bibnamefont {Cobden}}, \bibinfo {author} {\bibfnamefont {J.-H.}\
  \bibnamefont {Chu}},\ and\ \bibinfo {author} {\bibfnamefont {X.}~\bibnamefont
  {Xu}},\ }\bibfield  {title} {\bibinfo {title} {Two-dimensional itinerant
  ferromagnetism in atomically thin {$\mathrm{Fe_3GeTe_2}$}},\ }\href
  {https://doi.org/10.1038/s41563-018-0149-7} {\bibfield  {journal} {\bibinfo
  {journal} {Nature Materials}\ }\textbf {\bibinfo {volume} {17}},\ \bibinfo
  {pages} {778} (\bibinfo {year} {2018})}\BibitemShut {NoStop}%
\bibitem [{\citenamefont {Li}\ \emph {et~al.}(2023)\citenamefont {Li},
  \citenamefont {Haldar}, \citenamefont {Drevelow},\ and\ \citenamefont
  {Heinze}}]{Fe3T1}%
  \BibitemOpen
  \bibfield  {author} {\bibinfo {author} {\bibfnamefont {D.}~\bibnamefont
  {Li}}, \bibinfo {author} {\bibfnamefont {S.}~\bibnamefont {Haldar}}, \bibinfo
  {author} {\bibfnamefont {T.}~\bibnamefont {Drevelow}},\ and\ \bibinfo
  {author} {\bibfnamefont {S.}~\bibnamefont {Heinze}},\ }\bibfield  {title}
  {\bibinfo {title} {Tuning the magnetic interactions in van der waals
  {${\mathrm{Fe}}_{3}{\mathrm{GeTe}}_{2}$} heterostructures: A comparative
  study of ab initio methods},\ }\href
  {https://doi.org/10.1103/PhysRevB.107.104428} {\bibfield  {journal} {\bibinfo
   {journal} {Physical Review B}\ }\textbf {\bibinfo {volume} {107}},\ \bibinfo
  {pages} {104428} (\bibinfo {year} {2023})}\BibitemShut {NoStop}%
\bibitem [{\citenamefont {Zhuang}\ \emph {et~al.}(2016)\citenamefont {Zhuang},
  \citenamefont {Kent},\ and\ \citenamefont {Hennig}}]{Fe3T2}%
  \BibitemOpen
  \bibfield  {author} {\bibinfo {author} {\bibfnamefont {H.~L.}\ \bibnamefont
  {Zhuang}}, \bibinfo {author} {\bibfnamefont {P.~R.~C.}\ \bibnamefont
  {Kent}},\ and\ \bibinfo {author} {\bibfnamefont {R.~G.}\ \bibnamefont
  {Hennig}},\ }\bibfield  {title} {\bibinfo {title} {Strong anisotropy and
  magnetostriction in the two-dimensional stoner ferromagnet
  {${\mathrm{Fe}}_{3}{\mathrm{GeTe}}_{2}$}},\ }\href
  {https://doi.org/10.1103/PhysRevB.93.134407} {\bibfield  {journal} {\bibinfo
  {journal} {Physical Review B}\ }\textbf {\bibinfo {volume} {93}},\ \bibinfo
  {pages} {134407} (\bibinfo {year} {2016})}\BibitemShut {NoStop}%
\bibitem [{\citenamefont {Zhou}\ \emph {et~al.}(2023)\citenamefont {Zhou},
  \citenamefont {Huang}, \citenamefont {Tang}, \citenamefont {Qiu},
  \citenamefont {Qin}, \citenamefont {Zhang}, \citenamefont {Li}, \citenamefont
  {Wu},\ and\ \citenamefont {Yuan}}]{Fe3IM}%
  \BibitemOpen
  \bibfield  {author} {\bibinfo {author} {\bibfnamefont {L.}~\bibnamefont
  {Zhou}}, \bibinfo {author} {\bibfnamefont {J.}~\bibnamefont {Huang}},
  \bibinfo {author} {\bibfnamefont {M.}~\bibnamefont {Tang}}, \bibinfo {author}
  {\bibfnamefont {C.}~\bibnamefont {Qiu}}, \bibinfo {author} {\bibfnamefont
  {F.}~\bibnamefont {Qin}}, \bibinfo {author} {\bibfnamefont {C.}~\bibnamefont
  {Zhang}}, \bibinfo {author} {\bibfnamefont {Z.}~\bibnamefont {Li}}, \bibinfo
  {author} {\bibfnamefont {D.}~\bibnamefont {Wu}},\ and\ \bibinfo {author}
  {\bibfnamefont {H.}~\bibnamefont {Yuan}},\ }\bibfield  {title} {\bibinfo
  {title} {Gate-tunable spin valve effect in {$\mathrm{Fe_3GeTe_2}$}-based van
  der waals heterostructures},\ }\href
  {https://doi.org/https://doi.org/10.1002/inf2.12371} {\bibfield  {journal}
  {\bibinfo  {journal} {InfoMat}\ }\textbf {\bibinfo {volume} {5}},\ \bibinfo
  {pages} {e12371} (\bibinfo {year} {2023})}\BibitemShut {NoStop}%
\bibitem [{\citenamefont {Huang}\ \emph {et~al.}(2022)\citenamefont {Huang},
  \citenamefont {Yao}, \citenamefont {Qi}, \citenamefont {Shen},\ and\
  \citenamefont {Cao}}]{Fe5k}%
  \BibitemOpen
  \bibfield  {author} {\bibinfo {author} {\bibfnamefont {Y.}~\bibnamefont
  {Huang}}, \bibinfo {author} {\bibfnamefont {X.}~\bibnamefont {Yao}}, \bibinfo
  {author} {\bibfnamefont {F.}~\bibnamefont {Qi}}, \bibinfo {author}
  {\bibfnamefont {W.}~\bibnamefont {Shen}},\ and\ \bibinfo {author}
  {\bibfnamefont {G.}~\bibnamefont {Cao}},\ }\bibfield  {title} {\bibinfo
  {title} {Anomalous resistivity upturn in the van der waals ferromagnet
  {$\mathrm{Fe_5GeTe_2}$}},\ }\href {https://doi.org/10.1063/5.0109735}
  {\bibfield  {journal} {\bibinfo  {journal} {Applied Physics Letters}\
  }\textbf {\bibinfo {volume} {121}},\ \bibinfo {pages} {162403} (\bibinfo
  {year} {2022})}\BibitemShut {NoStop}%
\bibitem [{\citenamefont {Deng}\ \emph {et~al.}(2022)\citenamefont {Deng},
  \citenamefont {Xiang}, \citenamefont {Lei}, \citenamefont {Zhu},
  \citenamefont {Mu}, \citenamefont {Zhuo}, \citenamefont {Hua}, \citenamefont
  {Wang}, \citenamefont {Wang}, \citenamefont {Wang}, \citenamefont {Tian},\
  and\ \citenamefont {Chen}}]{Fe5a1}%
  \BibitemOpen
  \bibfield  {author} {\bibinfo {author} {\bibfnamefont {Y.}~\bibnamefont
  {Deng}}, \bibinfo {author} {\bibfnamefont {Z.}~\bibnamefont {Xiang}},
  \bibinfo {author} {\bibfnamefont {B.}~\bibnamefont {Lei}}, \bibinfo {author}
  {\bibfnamefont {K.}~\bibnamefont {Zhu}}, \bibinfo {author} {\bibfnamefont
  {H.}~\bibnamefont {Mu}}, \bibinfo {author} {\bibfnamefont {W.}~\bibnamefont
  {Zhuo}}, \bibinfo {author} {\bibfnamefont {X.}~\bibnamefont {Hua}}, \bibinfo
  {author} {\bibfnamefont {M.}~\bibnamefont {Wang}}, \bibinfo {author}
  {\bibfnamefont {Z.}~\bibnamefont {Wang}}, \bibinfo {author} {\bibfnamefont
  {G.}~\bibnamefont {Wang}}, \bibinfo {author} {\bibfnamefont {M.}~\bibnamefont
  {Tian}},\ and\ \bibinfo {author} {\bibfnamefont {X.}~\bibnamefont {Chen}},\
  }\bibfield  {title} {\bibinfo {title} {Layer-number-dependent magnetism and
  anomalous hall effect in van der waals ferromagnet {$\mathrm{Fe_5GeTe_2}$}},\
  }\href {https://doi.org/10.1021/acs.nanolett.2c02696} {\bibfield  {journal}
  {\bibinfo  {journal} {Nano Letters}\ }\textbf {\bibinfo {volume} {22}},\
  \bibinfo {pages} {9839} (\bibinfo {year} {2022})}\BibitemShut {NoStop}%
\bibitem [{\citenamefont {Yang}\ \emph {et~al.}(2021)\citenamefont {Yang},
  \citenamefont {Zhou}, \citenamefont {Feng},\ and\ \citenamefont
  {Yao}}]{Femo}%
  \BibitemOpen
  \bibfield  {author} {\bibinfo {author} {\bibfnamefont {X.}~\bibnamefont
  {Yang}}, \bibinfo {author} {\bibfnamefont {X.}~\bibnamefont {Zhou}}, \bibinfo
  {author} {\bibfnamefont {W.}~\bibnamefont {Feng}},\ and\ \bibinfo {author}
  {\bibfnamefont {Y.}~\bibnamefont {Yao}},\ }\bibfield  {title} {\bibinfo
  {title} {Strong magneto-optical effect and anomalous transport in the
  two-dimensional van der waals magnets
  {${\mathrm{Fe}}_{n}{\mathrm{GeTe}}_{2}$} ($n=3$, 4, 5)},\ }\href
  {https://doi.org/10.1103/PhysRevB.104.104427} {\bibfield  {journal} {\bibinfo
   {journal} {Physical Review B}\ }\textbf {\bibinfo {volume} {104}},\ \bibinfo
  {pages} {104427} (\bibinfo {year} {2021})}\BibitemShut {NoStop}%
\bibitem [{\citenamefont {Ohta}\ \emph {et~al.}(2021)\citenamefont {Ohta},
  \citenamefont {Tokuda}, \citenamefont {Iwakiri}, \citenamefont {Sakai},
  \citenamefont {Driesen}, \citenamefont {Okada}, \citenamefont {Kobayashi},\
  and\ \citenamefont {Niimi}}]{Fe5bm}%
  \BibitemOpen
  \bibfield  {author} {\bibinfo {author} {\bibfnamefont {T.}~\bibnamefont
  {Ohta}}, \bibinfo {author} {\bibfnamefont {M.}~\bibnamefont {Tokuda}},
  \bibinfo {author} {\bibfnamefont {S.}~\bibnamefont {Iwakiri}}, \bibinfo
  {author} {\bibfnamefont {K.}~\bibnamefont {Sakai}}, \bibinfo {author}
  {\bibfnamefont {B.}~\bibnamefont {Driesen}}, \bibinfo {author} {\bibfnamefont
  {Y.}~\bibnamefont {Okada}}, \bibinfo {author} {\bibfnamefont
  {K.}~\bibnamefont {Kobayashi}},\ and\ \bibinfo {author} {\bibfnamefont
  {Y.}~\bibnamefont {Niimi}},\ }\bibfield  {title} {\bibinfo {title}
  {Butterfly-shaped magnetoresistance in van der waals ferromagnet
  {$\mathrm{Fe_5GeTe_2}$}},\ }\href {https://doi.org/10.1063/9.0000067}
  {\bibfield  {journal} {\bibinfo  {journal} {AIP Advances}\ }\textbf {\bibinfo
  {volume} {11}},\ \bibinfo {pages} {025014} (\bibinfo {year}
  {2021})}\BibitemShut {NoStop}%
\bibitem [{\citenamefont {Lv}\ \emph {et~al.}(2022)\citenamefont {Lv},
  \citenamefont {Pei}, \citenamefont {Yang}, \citenamefont {Qin}, \citenamefont
  {Liu}, \citenamefont {Zhang},\ and\ \citenamefont {Che}}]{Fe5m4}%
  \BibitemOpen
  \bibfield  {author} {\bibinfo {author} {\bibfnamefont {X.}~\bibnamefont
  {Lv}}, \bibinfo {author} {\bibfnamefont {K.}~\bibnamefont {Pei}}, \bibinfo
  {author} {\bibfnamefont {C.}~\bibnamefont {Yang}}, \bibinfo {author}
  {\bibfnamefont {G.}~\bibnamefont {Qin}}, \bibinfo {author} {\bibfnamefont
  {M.}~\bibnamefont {Liu}}, \bibinfo {author} {\bibfnamefont {J.}~\bibnamefont
  {Zhang}},\ and\ \bibinfo {author} {\bibfnamefont {R.}~\bibnamefont {Che}},\
  }\bibfield  {title} {\bibinfo {title} {Controllable topological magnetic
  transformations in the thickness-tunable van der waals ferromagnet
  {$\mathrm{Fe_5GeTe_2}$}},\ }\href {https://doi.org/10.1021/acsnano.2c08844}
  {\bibfield  {journal} {\bibinfo  {journal} {ACS Nano}\ }\textbf {\bibinfo
  {volume} {16}},\ \bibinfo {pages} {19319} (\bibinfo {year}
  {2022})}\BibitemShut {NoStop}%
\bibitem [{\citenamefont {Zhang}\ \emph {et~al.}(2022)\citenamefont {Zhang},
  \citenamefont {Liu}, \citenamefont {Zhang}, \citenamefont {Zhou},
  \citenamefont {Guan}, \citenamefont {Ma}, \citenamefont {Algaidi},
  \citenamefont {Zheng}, \citenamefont {Li}, \citenamefont {He}, \citenamefont
  {Zhang}, \citenamefont {Li}, \citenamefont {Hou}, \citenamefont {Yin},
  \citenamefont {Liu}, \citenamefont {Peng},\ and\ \citenamefont
  {Zhang}}]{Fems}%
  \BibitemOpen
  \bibfield  {author} {\bibinfo {author} {\bibfnamefont {C.}~\bibnamefont
  {Zhang}}, \bibinfo {author} {\bibfnamefont {C.}~\bibnamefont {Liu}}, \bibinfo
  {author} {\bibfnamefont {S.}~\bibnamefont {Zhang}}, \bibinfo {author}
  {\bibfnamefont {B.}~\bibnamefont {Zhou}}, \bibinfo {author} {\bibfnamefont
  {C.}~\bibnamefont {Guan}}, \bibinfo {author} {\bibfnamefont {Y.}~\bibnamefont
  {Ma}}, \bibinfo {author} {\bibfnamefont {H.}~\bibnamefont {Algaidi}},
  \bibinfo {author} {\bibfnamefont {D.}~\bibnamefont {Zheng}}, \bibinfo
  {author} {\bibfnamefont {Y.}~\bibnamefont {Li}}, \bibinfo {author}
  {\bibfnamefont {X.}~\bibnamefont {He}}, \bibinfo {author} {\bibfnamefont
  {J.}~\bibnamefont {Zhang}}, \bibinfo {author} {\bibfnamefont
  {P.}~\bibnamefont {Li}}, \bibinfo {author} {\bibfnamefont {Z.}~\bibnamefont
  {Hou}}, \bibinfo {author} {\bibfnamefont {G.}~\bibnamefont {Yin}}, \bibinfo
  {author} {\bibfnamefont {K.}~\bibnamefont {Liu}}, \bibinfo {author}
  {\bibfnamefont {Y.}~\bibnamefont {Peng}},\ and\ \bibinfo {author}
  {\bibfnamefont {X.-X.}\ \bibnamefont {Zhang}},\ }\bibfield  {title} {\bibinfo
  {title} {Magnetic skyrmions with unconventional helicity polarization in a
  van der waals ferromagnet},\ }\href
  {https://doi.org/https://doi.org/10.1002/adma.202204163} {\bibfield
  {journal} {\bibinfo  {journal} {Advanced Materials}\ }\textbf {\bibinfo
  {volume} {34}},\ \bibinfo {pages} {2204163} (\bibinfo {year}
  {2022})}\BibitemShut {NoStop}%
\bibitem [{\citenamefont {Schmitt}\ \emph {et~al.}(2022)\citenamefont
  {Schmitt}, \citenamefont {Denneulin}, \citenamefont {Kovács}, \citenamefont
  {Saunderson}, \citenamefont {Rüßmann}, \citenamefont {Shahee},
  \citenamefont {Scholz}, \citenamefont {Tavabi}, \citenamefont {Gradhand},
  \citenamefont {Mavropoulos}, \citenamefont {Lotsch}, \citenamefont
  {Dunin-Borkowski}, \citenamefont {Mokrousov}, \citenamefont {Blügel},\ and\
  \citenamefont {Kläui}}]{Fe5s1}%
  \BibitemOpen
  \bibfield  {author} {\bibinfo {author} {\bibfnamefont {M.}~\bibnamefont
  {Schmitt}}, \bibinfo {author} {\bibfnamefont {T.}~\bibnamefont {Denneulin}},
  \bibinfo {author} {\bibfnamefont {A.}~\bibnamefont {Kovács}}, \bibinfo
  {author} {\bibfnamefont {T.~G.}\ \bibnamefont {Saunderson}}, \bibinfo
  {author} {\bibfnamefont {P.}~\bibnamefont {Rüßmann}}, \bibinfo {author}
  {\bibfnamefont {A.}~\bibnamefont {Shahee}}, \bibinfo {author} {\bibfnamefont
  {T.}~\bibnamefont {Scholz}}, \bibinfo {author} {\bibfnamefont {A.~H.}\
  \bibnamefont {Tavabi}}, \bibinfo {author} {\bibfnamefont {M.}~\bibnamefont
  {Gradhand}}, \bibinfo {author} {\bibfnamefont {P.}~\bibnamefont
  {Mavropoulos}}, \bibinfo {author} {\bibfnamefont {B.~V.}\ \bibnamefont
  {Lotsch}}, \bibinfo {author} {\bibfnamefont {R.~E.}\ \bibnamefont
  {Dunin-Borkowski}}, \bibinfo {author} {\bibfnamefont {Y.}~\bibnamefont
  {Mokrousov}}, \bibinfo {author} {\bibfnamefont {S.}~\bibnamefont {Blügel}},\
  and\ \bibinfo {author} {\bibfnamefont {M.}~\bibnamefont {Kläui}},\
  }\bibfield  {title} {\bibinfo {title} {Skyrmionic spin structures in layered
  {$\mathrm{Fe_5GeTe_2}$} up to room temperature},\ }\href
  {https://doi.org/10.1038/s42005-022-01031-w} {\bibfield  {journal} {\bibinfo
  {journal} {Communications Physics}\ }\textbf {\bibinfo {volume} {5}},\
  \bibinfo {pages} {254} (\bibinfo {year} {2022})}\BibitemShut {NoStop}%
\bibitem [{\citenamefont {Sutherland}(1986)}]{dice0}%
  \BibitemOpen
  \bibfield  {author} {\bibinfo {author} {\bibfnamefont {B.}~\bibnamefont
  {Sutherland}},\ }\bibfield  {title} {\bibinfo {title} {Localization of
  electronic wave functions due to local topology},\ }\href
  {https://doi.org/10.1103/PhysRevB.34.5208} {\bibfield  {journal} {\bibinfo
  {journal} {Physical Review B}\ }\textbf {\bibinfo {volume} {34}},\ \bibinfo
  {pages} {5208} (\bibinfo {year} {1986})}\BibitemShut {NoStop}%
\bibitem [{\citenamefont {Călugăru}\ \emph {et~al.}(2021)\citenamefont
  {Călugăru}, \citenamefont {Chew}, \citenamefont {Elcoro}, \citenamefont
  {Xu}, \citenamefont {Regnault}, \citenamefont {Song},\ and\ \citenamefont
  {Bernevig}}]{FB1}%
  \BibitemOpen
  \bibfield  {author} {\bibinfo {author} {\bibfnamefont {D.}~\bibnamefont
  {Călugăru}}, \bibinfo {author} {\bibfnamefont {A.}~\bibnamefont {Chew}},
  \bibinfo {author} {\bibfnamefont {L.}~\bibnamefont {Elcoro}}, \bibinfo
  {author} {\bibfnamefont {Y.}~\bibnamefont {Xu}}, \bibinfo {author}
  {\bibfnamefont {N.}~\bibnamefont {Regnault}}, \bibinfo {author}
  {\bibfnamefont {Z.-D.}\ \bibnamefont {Song}},\ and\ \bibinfo {author}
  {\bibfnamefont {B.~A.}\ \bibnamefont {Bernevig}},\ }\bibfield  {title}
  {\bibinfo {title} {General construction and topological classification of
  crystalline flat bands},\ }\href {https://doi.org/10.1038/s41567-021-01445-3}
  {\bibfield  {journal} {\bibinfo  {journal} {Nature Physics}\ }\textbf
  {\bibinfo {volume} {18}},\ \bibinfo {pages} {185} (\bibinfo {year}
  {2021})}\BibitemShut {NoStop}%
\bibitem [{\citenamefont {Wu}\ \emph {et~al.}(2023)\citenamefont {Wu},
  \citenamefont {Fang}, \citenamefont {Yang}, \citenamefont {Liu},
  \citenamefont {Peng}, \citenamefont {Li}, \citenamefont {Lin}, \citenamefont
  {Shi}, \citenamefont {Yang}, \citenamefont {Luo}, \citenamefont {Wang},
  \citenamefont {Yang}, \citenamefont {Lu},\ and\ \citenamefont {Du}}]{Fe45}%
  \BibitemOpen
  \bibfield  {author} {\bibinfo {author} {\bibfnamefont {B.}~\bibnamefont
  {Wu}}, \bibinfo {author} {\bibfnamefont {S.}~\bibnamefont {Fang}}, \bibinfo
  {author} {\bibfnamefont {J.}~\bibnamefont {Yang}}, \bibinfo {author}
  {\bibfnamefont {S.}~\bibnamefont {Liu}}, \bibinfo {author} {\bibfnamefont
  {Y.}~\bibnamefont {Peng}}, \bibinfo {author} {\bibfnamefont {Q.}~\bibnamefont
  {Li}}, \bibinfo {author} {\bibfnamefont {Z.}~\bibnamefont {Lin}}, \bibinfo
  {author} {\bibfnamefont {J.}~\bibnamefont {Shi}}, \bibinfo {author}
  {\bibfnamefont {W.}~\bibnamefont {Yang}}, \bibinfo {author} {\bibfnamefont
  {Z.}~\bibnamefont {Luo}}, \bibinfo {author} {\bibfnamefont {C.}~\bibnamefont
  {Wang}}, \bibinfo {author} {\bibfnamefont {J.}~\bibnamefont {Yang}}, \bibinfo
  {author} {\bibfnamefont {J.}~\bibnamefont {Lu}},\ and\ \bibinfo {author}
  {\bibfnamefont {H.}~\bibnamefont {Du}},\ }\bibfield  {title} {\bibinfo
  {title} {High-performance
  {${\mathrm{Fe}}_{x}{\mathrm{Ge}\mathrm{Te}}_{2}\text{\ensuremath{-}}$}based
  (x =$\phantom{\rule{0.1em}{0ex}}$4 or 5) van der waals magnetic tunnel
  junctions},\ }\href {https://doi.org/10.1103/PhysRevApplied.19.024037}
  {\bibfield  {journal} {\bibinfo  {journal} {Physical Review Applied}\
  }\textbf {\bibinfo {volume} {19}},\ \bibinfo {pages} {024037} (\bibinfo
  {year} {2023})}\BibitemShut {NoStop}%
\bibitem [{\citenamefont {Kim}\ \emph {et~al.}(2021)\citenamefont {Kim},
  \citenamefont {Lee}, \citenamefont {Jang}, \citenamefont {Kim},\ and\
  \citenamefont {Shim}}]{Fe4b}%
  \BibitemOpen
  \bibfield  {author} {\bibinfo {author} {\bibfnamefont {D.}~\bibnamefont
  {Kim}}, \bibinfo {author} {\bibfnamefont {C.}~\bibnamefont {Lee}}, \bibinfo
  {author} {\bibfnamefont {B.~G.}\ \bibnamefont {Jang}}, \bibinfo {author}
  {\bibfnamefont {K.}~\bibnamefont {Kim}},\ and\ \bibinfo {author}
  {\bibfnamefont {J.~H.}\ \bibnamefont {Shim}},\ }\bibfield  {title} {\bibinfo
  {title} {Drastic change of magnetic anisotropy in {$\mathrm{Fe_3GeTe_2}$} and
  {$\mathrm{Fe_4GeTe_2}$} monolayers under electric field studied by density
  functional theory},\ }\href {https://doi.org/10.1038/s41598-021-96639-3}
  {\bibfield  {journal} {\bibinfo  {journal} {Scientific Reports}\ }\textbf
  {\bibinfo {volume} {11}},\ \bibinfo {pages} {17567} (\bibinfo {year}
  {2021})}\BibitemShut {NoStop}%
\bibitem [{\citenamefont {Joe}\ \emph {et~al.}(2019)\citenamefont {Joe},
  \citenamefont {Yang},\ and\ \citenamefont {Lee}}]{Fe5b}%
  \BibitemOpen
  \bibfield  {author} {\bibinfo {author} {\bibfnamefont {M.}~\bibnamefont
  {Joe}}, \bibinfo {author} {\bibfnamefont {U.}~\bibnamefont {Yang}},\ and\
  \bibinfo {author} {\bibfnamefont {C.}~\bibnamefont {Lee}},\ }\bibfield
  {title} {\bibinfo {title} {First-principles study of ferromagnetic metal
  {$\mathrm{Fe_5GeTe_2}$}},\ }\href
  {https://doi.org/https://doi.org/10.1016/j.nanoms.2019.09.009} {\bibfield
  {journal} {\bibinfo  {journal} {Nano Materials Science}\ }\textbf {\bibinfo
  {volume} {1}},\ \bibinfo {pages} {299} (\bibinfo {year} {2019})}\BibitemShut
  {NoStop}%
\bibitem [{\citenamefont {Yamagami}\ \emph {et~al.}(2021)\citenamefont
  {Yamagami}, \citenamefont {Fujisawa}, \citenamefont {Driesen}, \citenamefont
  {Hsu}, \citenamefont {Kawaguchi}, \citenamefont {Tanaka}, \citenamefont
  {Kondo}, \citenamefont {Zhang}, \citenamefont {Wadati}, \citenamefont
  {Araki}, \citenamefont {Takeda}, \citenamefont {Takeda}, \citenamefont
  {Muro}, \citenamefont {Chuang}, \citenamefont {Niimi}, \citenamefont
  {Kuroda}, \citenamefont {Kobayashi},\ and\ \citenamefont {Okada}}]{Fe5b2}%
  \BibitemOpen
  \bibfield  {author} {\bibinfo {author} {\bibfnamefont {K.}~\bibnamefont
  {Yamagami}}, \bibinfo {author} {\bibfnamefont {Y.}~\bibnamefont {Fujisawa}},
  \bibinfo {author} {\bibfnamefont {B.}~\bibnamefont {Driesen}}, \bibinfo
  {author} {\bibfnamefont {C.~H.}\ \bibnamefont {Hsu}}, \bibinfo {author}
  {\bibfnamefont {K.}~\bibnamefont {Kawaguchi}}, \bibinfo {author}
  {\bibfnamefont {H.}~\bibnamefont {Tanaka}}, \bibinfo {author} {\bibfnamefont
  {T.}~\bibnamefont {Kondo}}, \bibinfo {author} {\bibfnamefont
  {Y.}~\bibnamefont {Zhang}}, \bibinfo {author} {\bibfnamefont
  {H.}~\bibnamefont {Wadati}}, \bibinfo {author} {\bibfnamefont
  {K.}~\bibnamefont {Araki}}, \bibinfo {author} {\bibfnamefont
  {T.}~\bibnamefont {Takeda}}, \bibinfo {author} {\bibfnamefont
  {Y.}~\bibnamefont {Takeda}}, \bibinfo {author} {\bibfnamefont
  {T.}~\bibnamefont {Muro}}, \bibinfo {author} {\bibfnamefont {F.~C.}\
  \bibnamefont {Chuang}}, \bibinfo {author} {\bibfnamefont {Y.}~\bibnamefont
  {Niimi}}, \bibinfo {author} {\bibfnamefont {K.}~\bibnamefont {Kuroda}},
  \bibinfo {author} {\bibfnamefont {M.}~\bibnamefont {Kobayashi}},\ and\
  \bibinfo {author} {\bibfnamefont {Y.}~\bibnamefont {Okada}},\ }\bibfield
  {title} {\bibinfo {title} {Itinerant ferromagnetism mediated by giant spin
  polarization of the metallic ligand band in the van der waals magnet
  {${\mathrm{Fe}}_{5}\mathrm{Ge}{\mathrm{Te}}_{2}$}},\ }\href
  {https://doi.org/10.1103/PhysRevB.103.L060403} {\bibfield  {journal}
  {\bibinfo  {journal} {Physical Review B}\ }\textbf {\bibinfo {volume}
  {103}},\ \bibinfo {pages} {L060403} (\bibinfo {year} {2021})}\BibitemShut
  {NoStop}%
\bibitem [{\citenamefont {Lieb}(1989)}]{FBL}%
  \BibitemOpen
  \bibfield  {author} {\bibinfo {author} {\bibfnamefont {E.~H.}\ \bibnamefont
  {Lieb}},\ }\bibfield  {title} {\bibinfo {title} {Two theorems on the hubbard
  model},\ }\href {https://doi.org/10.1103/PhysRevLett.62.1201} {\bibfield
  {journal} {\bibinfo  {journal} {Phys. Rev. Lett.}\ }\textbf {\bibinfo
  {volume} {62}},\ \bibinfo {pages} {1201} (\bibinfo {year}
  {1989})}\BibitemShut {NoStop}%
\bibitem [{\citenamefont {Derzhko}\ \emph {et~al.}(2015)\citenamefont
  {Derzhko}, \citenamefont {Richter},\ and\ \citenamefont {Maksymenko}}]{FBM}%
  \BibitemOpen
  \bibfield  {author} {\bibinfo {author} {\bibfnamefont {O.}~\bibnamefont
  {Derzhko}}, \bibinfo {author} {\bibfnamefont {J.}~\bibnamefont {Richter}},\
  and\ \bibinfo {author} {\bibfnamefont {M.}~\bibnamefont {Maksymenko}},\
  }\bibfield  {title} {\bibinfo {title} {Strongly correlated flat-band systems:
  The route from heisenberg spins to hubbard electrons},\ }\href
  {https://doi.org/10.1142/s0217979215300078} {\bibfield  {journal} {\bibinfo
  {journal} {International Journal of Modern Physics B}\ }\textbf {\bibinfo
  {volume} {29}},\ \bibinfo {pages} {1530007} (\bibinfo {year}
  {2015})}\BibitemShut {NoStop}%
\bibitem [{\citenamefont {Mielke}(1991)}]{FBM1}%
  \BibitemOpen
  \bibfield  {author} {\bibinfo {author} {\bibfnamefont {A.}~\bibnamefont
  {Mielke}},\ }\bibfield  {title} {\bibinfo {title} {Ferromagnetism in the
  hubbard model on line graphs and further considerations},\ }\href
  {https://doi.org/10.1088/0305-4470/24/14/018} {\bibfield  {journal} {\bibinfo
   {journal} {Journal of Physics A: Mathematical and General}\ }\textbf
  {\bibinfo {volume} {24}},\ \bibinfo {pages} {3311} (\bibinfo {year}
  {1991})}\BibitemShut {NoStop}%
\bibitem [{\citenamefont {Mielke}(1993{\natexlab{a}})}]{FBM2}%
  \BibitemOpen
  \bibfield  {author} {\bibinfo {author} {\bibfnamefont {A.}~\bibnamefont
  {Mielke}},\ }\bibfield  {title} {\bibinfo {title} {Ferromagnetism in the
  hubbard model and hund's rule},\ }\href
  {https://doi.org/https://doi.org/10.1016/0375-9601(93)90207-G} {\bibfield
  {journal} {\bibinfo  {journal} {Physics Letters A}\ }\textbf {\bibinfo
  {volume} {174}},\ \bibinfo {pages} {443} (\bibinfo {year}
  {1993}{\natexlab{a}})}\BibitemShut {NoStop}%
\bibitem [{\citenamefont {Tasaki}(1998)}]{FBM4}%
  \BibitemOpen
  \bibfield  {author} {\bibinfo {author} {\bibfnamefont {H.}~\bibnamefont
  {Tasaki}},\ }\bibfield  {title} {\bibinfo {title} {From nagaoka's
  ferromagnetism to flat-band ferromagnetism and beyond: An introduction to
  ferromagnetism in the hubbard model},\ }\href
  {https://doi.org/10.1143/ptp.99.489} {\bibfield  {journal} {\bibinfo
  {journal} {Progress of Theoretical Physics}\ }\textbf {\bibinfo {volume}
  {99}},\ \bibinfo {pages} {489} (\bibinfo {year} {1998})}\BibitemShut
  {NoStop}%
\bibitem [{\citenamefont {Tasaki}(1996)}]{FBM5}%
  \BibitemOpen
  \bibfield  {author} {\bibinfo {author} {\bibfnamefont {H.}~\bibnamefont
  {Tasaki}},\ }\bibfield  {title} {\bibinfo {title} {Stability of
  ferromagnetism in hubbard models with nearly flat bands},\ }\href
  {https://doi.org/10.1007/bf02179652} {\bibfield  {journal} {\bibinfo
  {journal} {Journal of Statistical Physics}\ }\textbf {\bibinfo {volume}
  {84}},\ \bibinfo {pages} {535} (\bibinfo {year} {1996})}\BibitemShut
  {NoStop}%
\bibitem [{\citenamefont {Kusakabe}\ and\ \citenamefont {Aoki}(1994)}]{FB2F}%
  \BibitemOpen
  \bibfield  {author} {\bibinfo {author} {\bibfnamefont {K.}~\bibnamefont
  {Kusakabe}}\ and\ \bibinfo {author} {\bibfnamefont {H.}~\bibnamefont
  {Aoki}},\ }\bibfield  {title} {\bibinfo {title} {Ferromagnetic spin-wave
  theory in the multiband hubbard model having a flat band},\ }\href
  {https://doi.org/10.1103/PhysRevLett.72.144} {\bibfield  {journal} {\bibinfo
  {journal} {Physical Review Letters}\ }\textbf {\bibinfo {volume} {72}},\
  \bibinfo {pages} {144} (\bibinfo {year} {1994})}\BibitemShut {NoStop}%
\bibitem [{\citenamefont {Kresse}\ and\ \citenamefont {Joubert}(1999)}]{vasp4}%
  \BibitemOpen
  \bibfield  {author} {\bibinfo {author} {\bibfnamefont {G.}~\bibnamefont
  {Kresse}}\ and\ \bibinfo {author} {\bibfnamefont {D.}~\bibnamefont
  {Joubert}},\ }\bibfield  {title} {\bibinfo {title} {From ultrasoft
  pseudopotentials to the projector augmented-wave method},\ }\href
  {https://doi.org/10.1103/PhysRevB.59.1758} {\bibfield  {journal} {\bibinfo
  {journal} {Physical Review B}\ }\textbf {\bibinfo {volume} {59}},\ \bibinfo
  {pages} {1758} (\bibinfo {year} {1999})}\BibitemShut {NoStop}%
\bibitem [{\citenamefont {Kresse}\ and\ \citenamefont
  {Furthmuller}(1996{\natexlab{a}})}]{vasp1}%
  \BibitemOpen
  \bibfield  {author} {\bibinfo {author} {\bibfnamefont {G.}~\bibnamefont
  {Kresse}}\ and\ \bibinfo {author} {\bibfnamefont {J.}~\bibnamefont
  {Furthmuller}},\ }\bibfield  {title} {\bibinfo {title} {Efficient iterative
  schemes for ab initio total-energy calculations using a plane-wave basis
  set},\ }\href {https://doi.org/10.1103/physrevb.54.11169} {\bibfield
  {journal} {\bibinfo  {journal} {Physical Review B}\ }\textbf {\bibinfo
  {volume} {54}},\ \bibinfo {pages} {11169} (\bibinfo {year}
  {1996}{\natexlab{a}})}\BibitemShut {NoStop}%
\bibitem [{\citenamefont {Kresse}\ and\ \citenamefont
  {Furthmuller}(1996{\natexlab{b}})}]{vasp2}%
  \BibitemOpen
  \bibfield  {author} {\bibinfo {author} {\bibfnamefont {G.}~\bibnamefont
  {Kresse}}\ and\ \bibinfo {author} {\bibfnamefont {J.}~\bibnamefont
  {Furthmuller}},\ }\bibfield  {title} {\bibinfo {title} {Efficiency of
  ab-initio total energy calculations for metals and semiconductors using a
  plane-wave basis set},\ }\href
  {https://doi.org/https://doi.org/10.1016/0927-0256(96)00008-0} {\bibfield
  {journal} {\bibinfo  {journal} {Computational Materials Science}\ }\textbf
  {\bibinfo {volume} {6}},\ \bibinfo {pages} {15} (\bibinfo {year}
  {1996}{\natexlab{b}})}\BibitemShut {NoStop}%
\bibitem [{\citenamefont {Kresse}\ \emph {et~al.}(1994)\citenamefont {Kresse},
  \citenamefont {Furthmuller},\ and\ \citenamefont {Hafner}}]{vasp3}%
  \BibitemOpen
  \bibfield  {author} {\bibinfo {author} {\bibfnamefont {G.}~\bibnamefont
  {Kresse}}, \bibinfo {author} {\bibfnamefont {J.}~\bibnamefont
  {Furthmuller}},\ and\ \bibinfo {author} {\bibfnamefont {J.}~\bibnamefont
  {Hafner}},\ }\bibfield  {title} {\bibinfo {title} {Theory of the crystal
  structures of selenium and tellurium: The effect of generalized-gradient
  corrections to the local-density approximation},\ }\href
  {https://doi.org/10.1103/PhysRevB.50.13181} {\bibfield  {journal} {\bibinfo
  {journal} {Physical Review B}\ }\textbf {\bibinfo {volume} {50}},\ \bibinfo
  {pages} {13181} (\bibinfo {year} {1994})}\BibitemShut {NoStop}%
\bibitem [{\citenamefont {Pizzi}\ \emph {et~al.}(2020)\citenamefont {Pizzi},
  \citenamefont {Vitale}, \citenamefont {Arita}, \citenamefont {Blügel},
  \citenamefont {Freimuth}, \citenamefont {Géranton}, \citenamefont
  {Gibertini}, \citenamefont {Gresch}, \citenamefont {Johnson}, \citenamefont
  {Koretsune}, \citenamefont {Ibañez-Azpiroz}, \citenamefont {Lee},
  \citenamefont {Lihm}, \citenamefont {Marchand}, \citenamefont {Marrazzo},
  \citenamefont {Mokrousov}, \citenamefont {Mustafa}, \citenamefont {Nohara},
  \citenamefont {Nomura}, \citenamefont {Paulatto}, \citenamefont {Poncé},
  \citenamefont {Ponweiser}, \citenamefont {Qiao}, \citenamefont {Thöle},
  \citenamefont {Tsirkin}, \citenamefont {Wierzbowska}, \citenamefont
  {Marzari}, \citenamefont {Vanderbilt}, \citenamefont {Souza}, \citenamefont
  {Mostofi},\ and\ \citenamefont {Yates}}]{W90}%
  \BibitemOpen
  \bibfield  {author} {\bibinfo {author} {\bibfnamefont {G.}~\bibnamefont
  {Pizzi}}, \bibinfo {author} {\bibfnamefont {V.}~\bibnamefont {Vitale}},
  \bibinfo {author} {\bibfnamefont {R.}~\bibnamefont {Arita}}, \bibinfo
  {author} {\bibfnamefont {S.}~\bibnamefont {Blügel}}, \bibinfo {author}
  {\bibfnamefont {F.}~\bibnamefont {Freimuth}}, \bibinfo {author}
  {\bibfnamefont {G.}~\bibnamefont {Géranton}}, \bibinfo {author}
  {\bibfnamefont {M.}~\bibnamefont {Gibertini}}, \bibinfo {author}
  {\bibfnamefont {D.}~\bibnamefont {Gresch}}, \bibinfo {author} {\bibfnamefont
  {C.}~\bibnamefont {Johnson}}, \bibinfo {author} {\bibfnamefont
  {T.}~\bibnamefont {Koretsune}}, \bibinfo {author} {\bibfnamefont
  {J.}~\bibnamefont {Ibañez-Azpiroz}}, \bibinfo {author} {\bibfnamefont
  {H.}~\bibnamefont {Lee}}, \bibinfo {author} {\bibfnamefont {J.-M.}\
  \bibnamefont {Lihm}}, \bibinfo {author} {\bibfnamefont {D.}~\bibnamefont
  {Marchand}}, \bibinfo {author} {\bibfnamefont {A.}~\bibnamefont {Marrazzo}},
  \bibinfo {author} {\bibfnamefont {Y.}~\bibnamefont {Mokrousov}}, \bibinfo
  {author} {\bibfnamefont {J.~I.}\ \bibnamefont {Mustafa}}, \bibinfo {author}
  {\bibfnamefont {Y.}~\bibnamefont {Nohara}}, \bibinfo {author} {\bibfnamefont
  {Y.}~\bibnamefont {Nomura}}, \bibinfo {author} {\bibfnamefont
  {L.}~\bibnamefont {Paulatto}}, \bibinfo {author} {\bibfnamefont
  {S.}~\bibnamefont {Poncé}}, \bibinfo {author} {\bibfnamefont
  {T.}~\bibnamefont {Ponweiser}}, \bibinfo {author} {\bibfnamefont
  {J.}~\bibnamefont {Qiao}}, \bibinfo {author} {\bibfnamefont {F.}~\bibnamefont
  {Thöle}}, \bibinfo {author} {\bibfnamefont {S.~S.}\ \bibnamefont {Tsirkin}},
  \bibinfo {author} {\bibfnamefont {M.}~\bibnamefont {Wierzbowska}}, \bibinfo
  {author} {\bibfnamefont {N.}~\bibnamefont {Marzari}}, \bibinfo {author}
  {\bibfnamefont {D.}~\bibnamefont {Vanderbilt}}, \bibinfo {author}
  {\bibfnamefont {I.}~\bibnamefont {Souza}}, \bibinfo {author} {\bibfnamefont
  {A.~A.}\ \bibnamefont {Mostofi}},\ and\ \bibinfo {author} {\bibfnamefont
  {J.~R.}\ \bibnamefont {Yates}},\ }\bibfield  {title} {\bibinfo {title}
  {Wannier90 as a community code: new features and applications},\ }\href
  {https://doi.org/10.1088/1361-648X/ab51ff} {\bibfield  {journal} {\bibinfo
  {journal} {Journal of Physics: Condensed Matter}\ }\textbf {\bibinfo {volume}
  {32}},\ \bibinfo {pages} {165902} (\bibinfo {year} {2020})}\BibitemShut
  {NoStop}%
\bibitem [{\citenamefont {Marzari}\ and\ \citenamefont
  {Vanderbilt}(1997)}]{W902}%
  \BibitemOpen
  \bibfield  {author} {\bibinfo {author} {\bibfnamefont {N.}~\bibnamefont
  {Marzari}}\ and\ \bibinfo {author} {\bibfnamefont {D.}~\bibnamefont
  {Vanderbilt}},\ }\bibfield  {title} {\bibinfo {title} {Maximally localized
  generalized wannier functions for composite energy bands},\ }\href
  {https://doi.org/10.1103/PhysRevB.56.12847} {\bibfield  {journal} {\bibinfo
  {journal} {Physical Review B}\ }\textbf {\bibinfo {volume} {56}},\ \bibinfo
  {pages} {12847} (\bibinfo {year} {1997})}\BibitemShut {NoStop}%
\bibitem [{\citenamefont {Souza}\ \emph {et~al.}(2001)\citenamefont {Souza},
  \citenamefont {Marzari},\ and\ \citenamefont {Vanderbilt}}]{W903}%
  \BibitemOpen
  \bibfield  {author} {\bibinfo {author} {\bibfnamefont {I.}~\bibnamefont
  {Souza}}, \bibinfo {author} {\bibfnamefont {N.}~\bibnamefont {Marzari}},\
  and\ \bibinfo {author} {\bibfnamefont {D.}~\bibnamefont {Vanderbilt}},\
  }\bibfield  {title} {\bibinfo {title} {Maximally localized wannier functions
  for entangled energy bands},\ }\href
  {https://doi.org/10.1103/PhysRevB.65.035109} {\bibfield  {journal} {\bibinfo
  {journal} {Physical Review B}\ }\textbf {\bibinfo {volume} {65}},\ \bibinfo
  {pages} {035109} (\bibinfo {year} {2001})}\BibitemShut {NoStop}%
\bibitem [{\citenamefont {Leykam}\ \emph {et~al.}(2018)\citenamefont {Leykam},
  \citenamefont {Andreanov},\ and\ \citenamefont {Flach}}]{FB2}%
  \BibitemOpen
  \bibfield  {author} {\bibinfo {author} {\bibfnamefont {D.}~\bibnamefont
  {Leykam}}, \bibinfo {author} {\bibfnamefont {A.}~\bibnamefont {Andreanov}},\
  and\ \bibinfo {author} {\bibfnamefont {S.}~\bibnamefont {Flach}},\ }\bibfield
   {title} {\bibinfo {title} {Artificial flat band systems: from lattice models
  to experiments},\ }\href {https://doi.org/10.1080/23746149.2018.1473052}
  {\bibfield  {journal} {\bibinfo  {journal} {Advances in Physics: X}\ }\textbf
  {\bibinfo {volume} {3}},\ \bibinfo {pages} {1473052} (\bibinfo {year}
  {2018})}\BibitemShut {NoStop}%
\bibitem [{\citenamefont {Sukhachov}\ \emph
  {et~al.}(2023{\natexlab{a}})\citenamefont {Sukhachov}, \citenamefont
  {Oriekhov},\ and\ \citenamefont {Gorbar}}]{ar1}%
  \BibitemOpen
  \bibfield  {author} {\bibinfo {author} {\bibfnamefont {P.~O.}\ \bibnamefont
  {Sukhachov}}, \bibinfo {author} {\bibfnamefont {D.~O.}\ \bibnamefont
  {Oriekhov}},\ and\ \bibinfo {author} {\bibfnamefont {E.~V.}\ \bibnamefont
  {Gorbar}},\ }\href@noop {} {\bibinfo {title} {Stackings and effective models
  of bilayer dice lattices}} (\bibinfo {year} {2023}{\natexlab{a}}),\ \Eprint
  {https://arxiv.org/abs/2303.01452} {arXiv:2303.01452 [cond-mat.mes-hall]}
  \BibitemShut {NoStop}%
\bibitem [{\citenamefont {Sukhachov}\ \emph
  {et~al.}(2023{\natexlab{b}})\citenamefont {Sukhachov}, \citenamefont
  {Oriekhov},\ and\ \citenamefont {Gorbar}}]{ar2}%
  \BibitemOpen
  \bibfield  {author} {\bibinfo {author} {\bibfnamefont {P.~O.}\ \bibnamefont
  {Sukhachov}}, \bibinfo {author} {\bibfnamefont {D.~O.}\ \bibnamefont
  {Oriekhov}},\ and\ \bibinfo {author} {\bibfnamefont {E.~V.}\ \bibnamefont
  {Gorbar}},\ }\href@noop {} {\bibinfo {title} {Optical conductivity of bilayer
  dice lattices}} (\bibinfo {year} {2023}{\natexlab{b}}),\ \Eprint
  {https://arxiv.org/abs/2303.08258} {arXiv:2303.08258 [cond-mat.mes-hall]}
  \BibitemShut {NoStop}%
\bibitem [{\citenamefont {Regnault}\ \emph {et~al.}(2022)\citenamefont
  {Regnault}, \citenamefont {Xu}, \citenamefont {Li}, \citenamefont {Ma},
  \citenamefont {Jovanovic}, \citenamefont {Yazdani}, \citenamefont {Parkin},
  \citenamefont {Felser}, \citenamefont {Schoop}, \citenamefont {Ong},
  \citenamefont {Cava}, \citenamefont {Elcoro}, \citenamefont {Song},\ and\
  \citenamefont {Bernevig}}]{FB3}%
  \BibitemOpen
  \bibfield  {author} {\bibinfo {author} {\bibfnamefont {N.}~\bibnamefont
  {Regnault}}, \bibinfo {author} {\bibfnamefont {Y.}~\bibnamefont {Xu}},
  \bibinfo {author} {\bibfnamefont {M.-R.}\ \bibnamefont {Li}}, \bibinfo
  {author} {\bibfnamefont {D.-S.}\ \bibnamefont {Ma}}, \bibinfo {author}
  {\bibfnamefont {M.}~\bibnamefont {Jovanovic}}, \bibinfo {author}
  {\bibfnamefont {A.}~\bibnamefont {Yazdani}}, \bibinfo {author} {\bibfnamefont
  {S.~S.~P.}\ \bibnamefont {Parkin}}, \bibinfo {author} {\bibfnamefont
  {C.}~\bibnamefont {Felser}}, \bibinfo {author} {\bibfnamefont {L.~M.}\
  \bibnamefont {Schoop}}, \bibinfo {author} {\bibfnamefont {N.~P.}\
  \bibnamefont {Ong}}, \bibinfo {author} {\bibfnamefont {R.~J.}\ \bibnamefont
  {Cava}}, \bibinfo {author} {\bibfnamefont {L.}~\bibnamefont {Elcoro}},
  \bibinfo {author} {\bibfnamefont {Z.-D.}\ \bibnamefont {Song}},\ and\
  \bibinfo {author} {\bibfnamefont {B.~A.}\ \bibnamefont {Bernevig}},\
  }\bibfield  {title} {\bibinfo {title} {Catalogue of flat-band stoichiometric
  materials},\ }\href {https://doi.org/10.1038/s41586-022-04519-1} {\bibfield
  {journal} {\bibinfo  {journal} {Nature}\ }\textbf {\bibinfo {volume} {603}},\
  \bibinfo {pages} {824} (\bibinfo {year} {2022})}\BibitemShut {NoStop}%
\bibitem [{\citenamefont {Li}\ \emph {et~al.}(2018)\citenamefont {Li},
  \citenamefont {Zhuang}, \citenamefont {Wang}, \citenamefont {Feng},
  \citenamefont {Gao}, \citenamefont {Xu}, \citenamefont {Hao}, \citenamefont
  {Wang}, \citenamefont {Zhang}, \citenamefont {Wu}, \citenamefont {Dou},
  \citenamefont {Chen}, \citenamefont {Hu},\ and\ \citenamefont {Du}}]{FBk}%
  \BibitemOpen
  \bibfield  {author} {\bibinfo {author} {\bibfnamefont {Z.}~\bibnamefont
  {Li}}, \bibinfo {author} {\bibfnamefont {J.}~\bibnamefont {Zhuang}}, \bibinfo
  {author} {\bibfnamefont {L.}~\bibnamefont {Wang}}, \bibinfo {author}
  {\bibfnamefont {H.}~\bibnamefont {Feng}}, \bibinfo {author} {\bibfnamefont
  {Q.}~\bibnamefont {Gao}}, \bibinfo {author} {\bibfnamefont {X.}~\bibnamefont
  {Xu}}, \bibinfo {author} {\bibfnamefont {W.}~\bibnamefont {Hao}}, \bibinfo
  {author} {\bibfnamefont {X.}~\bibnamefont {Wang}}, \bibinfo {author}
  {\bibfnamefont {C.}~\bibnamefont {Zhang}}, \bibinfo {author} {\bibfnamefont
  {K.}~\bibnamefont {Wu}}, \bibinfo {author} {\bibfnamefont {S.~X.}\
  \bibnamefont {Dou}}, \bibinfo {author} {\bibfnamefont {L.}~\bibnamefont
  {Chen}}, \bibinfo {author} {\bibfnamefont {Z.}~\bibnamefont {Hu}},\ and\
  \bibinfo {author} {\bibfnamefont {Y.}~\bibnamefont {Du}},\ }\bibfield
  {title} {\bibinfo {title} {Realization of flat band with possible nontrivial
  topology in electronic kagome lattice},\ }\href
  {https://doi.org/doi:10.1126/sciadv.aau4511} {\bibfield  {journal} {\bibinfo
  {journal} {Science Advances}\ }\textbf {\bibinfo {volume} {4}},\ \bibinfo
  {pages} {eaau4511} (\bibinfo {year} {2018})}\BibitemShut {NoStop}%
\bibitem [{\citenamefont {Kübler}(2021)}]{Itm1}%
  \BibitemOpen
  \bibfield  {author} {\bibinfo {author} {\bibfnamefont {J.}~\bibnamefont
  {Kübler}},\ }\href {https://doi.org/10.1093/oso/9780192895639.001.0001}
  {\emph {\bibinfo {title} {Theory of Itinerant Electron Magnetism, 2nd
  Edition}}}\ (\bibinfo  {publisher} {Oxford University Press},\ \bibinfo
  {year} {2021})\BibitemShut {NoStop}%
\bibitem [{\citenamefont {Wahle}\ \emph {et~al.}(1998)\citenamefont {Wahle},
  \citenamefont {Bl\"umer}, \citenamefont {Schlipf}, \citenamefont {Held},\
  and\ \citenamefont {Vollhardt}}]{Itm2}%
  \BibitemOpen
  \bibfield  {author} {\bibinfo {author} {\bibfnamefont {J.}~\bibnamefont
  {Wahle}}, \bibinfo {author} {\bibfnamefont {N.}~\bibnamefont {Bl\"umer}},
  \bibinfo {author} {\bibfnamefont {J.}~\bibnamefont {Schlipf}}, \bibinfo
  {author} {\bibfnamefont {K.}~\bibnamefont {Held}},\ and\ \bibinfo {author}
  {\bibfnamefont {D.}~\bibnamefont {Vollhardt}},\ }\bibfield  {title} {\bibinfo
  {title} {Microscopic conditions favoring itinerant ferromagnetism},\ }\href
  {https://doi.org/10.1103/PhysRevB.58.12749} {\bibfield  {journal} {\bibinfo
  {journal} {Physical Review B}\ }\textbf {\bibinfo {volume} {58}},\ \bibinfo
  {pages} {12749} (\bibinfo {year} {1998})}\BibitemShut {NoStop}%
\bibitem [{\citenamefont {Mielke}(1993{\natexlab{b}})}]{Itm3}%
  \BibitemOpen
  \bibfield  {author} {\bibinfo {author} {\bibfnamefont {A.}~\bibnamefont
  {Mielke}},\ }\bibfield  {title} {\bibinfo {title} {Ferromagnetism in the
  hubbard model and hund's rule},\ }\href
  {https://doi.org/https://doi.org/10.1016/0375-9601(93)90207-G} {\bibfield
  {journal} {\bibinfo  {journal} {Physics Letters A}\ }\textbf {\bibinfo
  {volume} {174}},\ \bibinfo {pages} {443} (\bibinfo {year}
  {1993}{\natexlab{b}})}\BibitemShut {NoStop}%
\bibitem [{\citenamefont {May}\ \emph {et~al.}(2016)\citenamefont {May},
  \citenamefont {Calder}, \citenamefont {Cantoni}, \citenamefont {Cao},\ and\
  \citenamefont {McGuire}}]{Fe3nd}%
  \BibitemOpen
  \bibfield  {author} {\bibinfo {author} {\bibfnamefont {A.~F.}\ \bibnamefont
  {May}}, \bibinfo {author} {\bibfnamefont {S.}~\bibnamefont {Calder}},
  \bibinfo {author} {\bibfnamefont {C.}~\bibnamefont {Cantoni}}, \bibinfo
  {author} {\bibfnamefont {H.}~\bibnamefont {Cao}},\ and\ \bibinfo {author}
  {\bibfnamefont {M.~A.}\ \bibnamefont {McGuire}},\ }\bibfield  {title}
  {\bibinfo {title} {Magnetic structure and phase stability of the van der
  waals bonded ferromagnet {${\mathrm{Fe}}_{3-x}{\mathrm{GeTe}}_{2}$}},\ }\href
  {https://doi.org/10.1103/PhysRevB.93.014411} {\bibfield  {journal} {\bibinfo
  {journal} {Physical Review B}\ }\textbf {\bibinfo {volume} {93}},\ \bibinfo
  {pages} {014411} (\bibinfo {year} {2016})}\BibitemShut {NoStop}%
\end{thebibliography}%

\end{document}